\documentclass[fleqn,usenatbib]{mnras}

\usepackage{newtxtext,newtxmath}

\usepackage[T1]{fontenc}
\usepackage{ae,aecompl}

\usepackage{graphicx}	% Including figure files
\usepackage{amsmath}	% Advanced maths commands
\usepackage{amssymb}	% Extra maths symbols
\usepackage{multirow}
\usepackage{bbold}
\usepackage[dvipsnames]{xcolor}

\newcommand{\D}{\text{d}}

\newcommand{\x}{\textbf{\textit{x}}}

\newcommand{\y}{\textbf{y}}

\newcommand{\n}{\hat{\textbf{\textit{n}}}}

\newcommand{\T}{\text{T}}

\newcommand{\orcid}[1]{{\hskip.5mm \href{#1}{\includegraphics[height=8px]{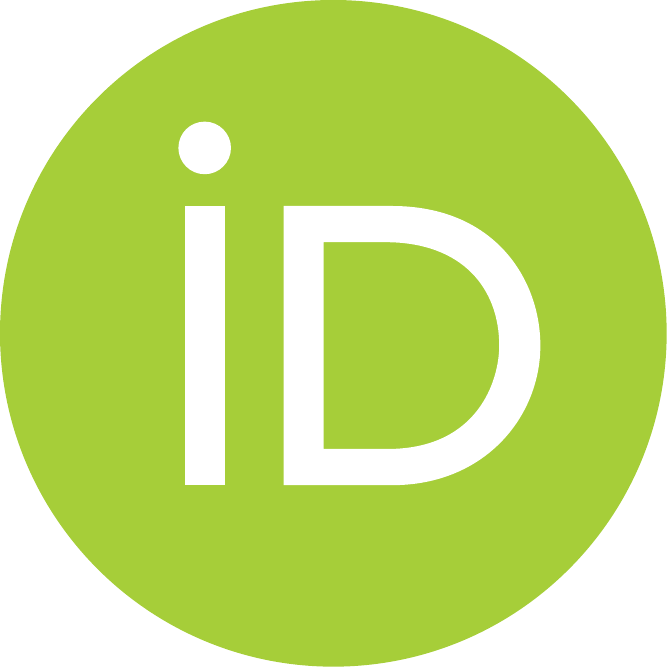}} \hskip.5mm}}

\newcommand{\bs}[1]{\boldsymbol{#1}}
\newcommand{\bm}[1]{\mathbfss{#1}}

\newcommand{\change}[1]{{\color{Fuchsia}#1}}

\usepackage{mathtools}

\usepackage[math]{cellspace}
\usepackage{cellspace}
\title
[Radiative Transfer as Regression]
{Radiative Transfer as a Bayesian Linear Regression problem}

\author
[F. De Ceuster et al.]
{F. De Ceuster$^{\orcid{https://orcid.org/0000-0001-5887-8498}, 1, 2, }$\thanks{Contact e-mail: \href{frederik.deceuster@kuleuven.be}{frederik.deceuster@kuleuven.be}.},
    T. Ceulemans$^{1}$,
    J. Cockayne$^{3}$,
    L. Decin$^{1, 4}$,
    and
    J. Yates$^{2}$
  \\ \\
  $^{1}$ Institute of Astronomy, KU Leuven, Celestijnenlaan 200D, 3001 Leuven, Belgium \\
  $^{2}$ Department of Physics and Astronomy, University College London, Gower Place, London, WC1E 6BT, UK \\
  $^{3}$ Department of Mathematical Sciences, University of Southampton, SO17 1BJ, UK \\
  $^{4}$ School of Chemistry, University of Leeds, Leeds LS2 9JT, UK \\
}

\date{Accepted XXX. Received YYY; in original form ZZZ}

\pubyear{2022}

\begin{document}
\label{firstpage}
\pagerange{\pageref{firstpage}--\pageref{lastpage}}
\maketitle

\begin{abstract}
%   This is a simple template for authors to write new MNRAS papers.
%   The abstract should briefly describe the aims, methods, and main results of the paper.
%   It should be a single paragraph not more than 250 words (200 words for Letters).
%   No references should appear in the abstract.
    Electromagnetic radiation plays a crucial role in various physical and chemical processes.
    Hence, almost all astrophysical simulations require some form of radiative transfer model.
    Despite many innovations in radiative transfer algorithms and their implementation, realistic radiative transfer models remain very computationally expensive, such that one often has to resort to approximate descriptions.
    The complexity of these models makes it difficult to assess the validity of any approximation and to quantify uncertainties on the model results.
    This impedes scientific rigour, in particular, when comparing models to observations, or when using their results as input for other models.
    We present a probabilistic numerical approach to address these issues by treating radiative transfer as a Bayesian linear regression problem.
    This allows us to model uncertainties on the input and output of the model with the variances of the associated probability distributions.
    Furthermore, this approach naturally allows us to create reduced-order radiative transfer models with a quantifiable accuracy. These are approximate solutions to exact radiative transfer models, in contrast to the exact solutions to approximate models that are often used.
    As a first demonstration, we derive a probabilistic version of the method of characteristics, a commonly-used technique to solve radiative transfer problems.
\end{abstract}


%%%%%%%%%%%%%%%%%%%%%%%%%%%%%%%%%%%%%%%%%%%%%%%%%%

%%%%%%%%%%%%%%%%% BODY OF PAPER %%%%%%%%%%%%%%%%%%

%%%%%%%%%%%%%%%%%%%%%%%%%%%%%%%%%%%
\section{Introduction}
\label{sec:introduction}
%%%%%%%%%%%%%%%%%%%%%%%%%%%%%%%%%%%
Light, or electromagnetic radiation in general, is a key component of our Universe.
Not only does it dictate what we can or cannot observe, it also has the ability to significantly affect numerous physical and chemical processes ranging from radiative heating, cooling, and pressure in hydrodynamics to various photo-reactions in chemistry.
As a result, almost every astrophysical simulation requires some form of radiative transfer model.

Over the years, many different schemes have been devised to model radiative transfer, ranging from probabilistic Monte Carlo methods \citep[see e.g.][and the references therein]{Noebauer2019}, to several types of formal solvers \citep[see e.g.][and the references therein]{DeCeuster2019, Kanschat2009}.
Despite many improvements in computational efficiency and the use of modern computer hardware, realistic radiative transfer models keep posing a formidable computational challenge.
Consequently, one often has to resort to approximate descriptions of the radiation field, such as flux-limited diffusion \citep[see e.g.][]{Moens2022}, or parametrised radiative heating and cooling functions \citep[see e.g.][]{Xia2018}, which are often used in hydrodynamics models, or semi-analytical descriptions of the photon flux, which are used in photo-chemistry modelling \citep[see e.g.][]{VandeSande2019}.
Each of these approximate descriptions has its underlying assumptions and limitations, and
as models become larger and more complex, it becomes increasingly difficult to properly assess their validity, or to gauge the potential impact of a certain approximation on the results.

Every approximation induces uncertainty on the model result.
These uncertainties can either be intrinsic to the model, for instance, due to the discretisation of a continuous variable, or are due to the propagation of uncertainties through the model, for instance, due to uncertainties in the radiative constants of the medium that are used in the model.
Currently, most radiative transfer models lack any form of uncertainty quantification.
This is a severe shortcoming that impedes scientific rigour, in particular, when comparing these seemingly exact model results to naturally noisy observations.
In addition, as observations reach ever higher spatial and spectral resolutions \citep[see e.g.][]{Decin2020}, the model uncertainties become ever more relevant. 
Moreover, with the dawn of a new era of emulated models \citep[see e.g.][]{deMijola2019, Holdship2021, Kasim2022}, in which algorithms are trained rather than programmed, and simulation data is used as \emph{ground truth}, it is crucial, more than ever, to properly understand the uncertainties associated with simulations.
Unfortunately, due to the computational complexity of radiative transfer and the requirement for many approximations it remains very challenging to provide proper uncertainty quantification for most astrophysical radiative transfer models.

There is, however, an approach that can answer both to the need for approximation, because of computational efficiency, and the need for uncertainty quantification, for scientific rigour.
Instead of starting from an approximate description, the idea is to start from a more complete description and approximate (or compress) it into a smaller, more tractable, model.
There are two key advantages to this approach: (i) the approximation can be tailored to the problem at hand, and (ii) the uncertainty induced by the approximation can be estimated by the information lost in compressing the model.

In \cite{DeCeuster2020}, it was already shown that typical radiative transfer simulations of 3D hydrodynamics models, can often be compressed by more than an order of magnitude in size, without significant loss of accuracy, using a heuristic re-meshing algorithm.
This shows that accurate radiative transfer approximations can be obtained by compressing more precise models.
It remains, however, to quantify the uncertainties induced by compressing the model and to have more rigorous means to guide the now only heuristic compression algorithm.

In this paper, we propose a novel approach to quantify uncertainties, based on ideas from \emph{probabilistic numerics}. Specifically, we introduce a new numerical method, referred to in the literature as a probabilistic numerical method \citep{Henning2015, Hennig2022}, whose output is a probability distribution over solutions to the problem.
The mean of this distribution coincides with a traditional solution, while the variance can be interpreted as an uncertainty or error measure.
The advantages of this probabilistic approach over existing methods are that:
(i) since the output contains an intrinsic description of the approximation error, that error can be controlled without the need to compute expensive and often conservative error measures, 
(ii) as a full probability distribution, the description of error is richer than standard error measures, which typically constitute worst-case bounds on a global (i.e. norm-wise) or local error, and 
(iii) since the approximation error is expressed in a probabilistic manner, it can naturally be combined with other sources of uncertainty to provide a unified description of all uncertainty in the solution, as will be demonstrated in Section \ref{subsec:bayesian-moc}.
These three points make probabilistic numerical methods particularly appropriate for modelling radiative transfer, given the dire need for reliable uncertainty estimates on a computationally expensive model in the presence of uncertainties on input quantities such as radiative constants.

In particular, we propose to treat radiative transfer as a Bayesian linear regression problem.
The radiation field is modelled as the expectation of a multivariate Gaussian probability distribution over possible solutions for the radiation field, conditioned on evaluations of the radiative transfer equation and boundary conditions.
As such, the variance of the conditioned distribution can be used as a measure of uncertainty on this result.
The computational complexity of the regression model can be controlled by the dimension of the feature space, i.e. the number of basis functions that is used.
This allows us to create approximate (reduced-order) radiative transfer models, by reducing the number of basis functions.

The idea to treat function approximation or the solution of operator equations as a regression problem is certainly not new and can be traced all the way back to \cite{Poincare1896}, as pointed out by \cite{Diaconis1988}.
More recently, motivated by \cite{Henning2015}, these ideas gained renewed interest and now form an active research domain in applied mathematics known as probabilistic numerics
\citep[see e.g.][]{Cockayne2019, Hennig2022}.
For a comprehensive recent history, see e.g. \cite{Oates2019}.
The specific idea to view the solution of operator equations as a Bayesian linear regression problem has been proposed several times, by several different authors, and in several different contexts \citep[e.g.][]{vandenBoogaart2001, Graepel2003, Cockayne2017}.
While, in the literature, this is usually derived from a Gaussian process, here, we take a slightly more general point of view.

The probabilistic numerical method presented here is closely related to finite element methods \citep[see also e.g.][]{Girolami2021}.
In fact, the method just gives a probabilistic interpretation to an otherwise classical collocation method \citep[see e.g.][]{Kansa1990a, Kansa1990b, Fasshauer1999}.
Finite element methods were introduced in the context of astrophysical radiative transfer by \cite{Dykema1996}, who applied it to the moments of the radiative transfer equation.
Since then, these methods have successfully been applied in several astrophysical contexts \citep[see e.g.][]{Meier1999, Richling2001, Korcakova2003}.
Due to their widespread use, especially in industry, there is a vast body of research dedicated to uncertainty quantification for these methods \citep[see e.g.][and the references therein]{Verfurth2013}.
Also in the astrophysical context, for instance, \cite{Richling2001} proposed an error measure on their finite element radiative transfer solver, which they furthermore used to adapt their discretisation.
The key difference between classical finite element methods and the method presented here, is the probabilistic interpretation of the results, i.e. here the solutions are (conditioned) probability distributions over the space of possible solutions, rather than a single solution.
This allows us to also take into account the uncertainties on the model input, and furthermore facilitates the use of our model both in forward and inverse modelling pipelines.

Alternatively, the method presented in this paper can be viewed as a linear (and hence analytically solvable) version of a physics-informed neural network method \citep[see e.g.][]{Lagaris1998, Lagaris2000, Raissi2019}.
This technique, inspired by machine learning, to solve, for instance operator equations, has already been successfully applied to radiative transfer problems, for example by \cite{Mishra2021}, who, furthermore, derived rigorous error bounds for their results \citep{Mishra2021b}.
The assumption of linearity makes our model much simpler than this, and allows us to obtain analytic solutions which can be used, for instance, to relate it directly to the commonly-used method of characteristics.

The structure of this paper is as follows.
In Section \ref{sec:methods}, we introduce Bayesian linear regression and show how it can be used to solve linear operator equations in a way that naturally allows for uncertainty quantification.
In Section \ref{sec:prob_rad_trans}, we apply this to radiative transfer. We show how reduced-order radiative transfer models can be obtained, and we derive a Bayesian version of the method of characteristics.
Section \ref{sec:discussion} concerns future research towards practical implementations, and we conclude with Section \ref{sec:conclusion}.

%%%%%%%%%%%%%%%%%%%%%%%%%%%%%%%%%%%
\section{Methods}
\label{sec:methods}
%%%%%%%%%%%%%%%%%%%%%%%%%%%%%%%%%%%
We present a probabilistic numerical method to solve linear operator equations by treating it as a (Bayesian) linear regression problem.
This idea has already been discussed at length in the literature \citep[see e.g.][]{vandenBoogaart2001, Graepel2003, Cockayne2017}. Nevertheless, we present it again, but in a slightly more general way, to demonstrate its full potential for astrophysical modelling, and radiative transfer in particular.
For a more comprehensive introduction, see e.g. \cite{Bishop2006}, \cite{Rasmussen2006}, or \cite{Hennig2022}.

%%%%%%%%%%%%%%%%%%%%%%%%%%%%%%
\subsection{Linear Regression}
%%%%%%%%%%%%%%%%%%%%%%%%%%%%%%
The aim of a linear regression model is to approximate (or fit) a function, $f$, with a linear combination of basis functions, $\phi_{i}$, based on data in the form of function evaluations, $\left(x_{d},y_{d} \equiv f(x_{d})\right)$.
In this paper, we only consider real functions, so all variables are always assumed to be real.
Given a set of $N_{\text{b}}$ basis functions, $\{\phi_{i}\}$, and a set of $N_{\text{d}}$ data points, $\{(x_{d}, y_{d})\}$, the approximation can either be expanded in terms of the basis function or in terms of the data, resulting respectively in the primal and dual formulation. 

%%%%%%%%%%%%%%%%%%%%%%%%%%%%%%%%%%
\subsubsection{Primal formulation}
%%%%%%%%%%%%%%%%%%%%%%%%%%%%%%%%%%
In the primal formulation, the approximation, $\tilde{f} (x)$, is modelled as a linear combination of the basis functions,
\begin{equation}
    \tilde{f} (x)
	\ = \
	\sum_{i=1}^{N_{\text{b}}} w_{i} \, \phi_{i}(x)
	\ \equiv \
	\bs{w}^{\T} \bs{\phi}(x),
\label{eq:fap1}
\end{equation}
where we defined the weight vector, $\bs{w}$, and basis function vector, $\bs{\phi}$.
Appropriate weights, $w_{i}$, can be found, for instance, by minimising the regularised mean squared error between the model and the data,
\begin{align}
	\text{RMSE}(\bs{w})
	\ &\equiv \
 	\sum_{d=1}^{N_{\text{d}}}
 	\frac{1}{\sigma_{d}^{2}}
	\left(
		\bs{w}^{\T} \bs{\phi}(x_{d}) - y_{d}
	\right)^{2}
	\ + \
	\sum_{i=1}^{N_{\text{b}}}
	\left(\frac{w_{i}}{\lambda_{i}}\right)^{2} .
\label{eq:MSE0}
\end{align}
The factors, $\sigma_{d}^{-2}$, weight the contributions of the different data points to the mean error, and are summarised in the diagonal matrix $\bs{\sigma}\equiv\text{diag}(\sigma_{d})$.
We also added a regularisation term, characterised by the diagonal matrix $\bs{\lambda}\equiv\text{diag}(\lambda_{i})$, which penalises the size of the components of the weight vectors. This will guarantee the existence of a unique solution, as we will see below.
If we define the design matrix, $\Phi_{di} \equiv \phi_{i}(x_{d})$, and the data vector pair, $(\bs{x}, \bs{y})$, equation (\ref{eq:MSE0}) can conveniently be rewritten as,
\begin{align}
	\text{RMSE}(\bs{w})
	\ &\equiv \
	\left(
	    \bs{\sigma}^{-1}
	    \left(
	        \bs{\Phi} \bs{w}
	        -
	        \bs{y}
	    \right)
	\right)^{2}
	+
	\left(
	    \bs{\lambda}^{-1} \bs{w}
	\right)^{2} ,
\label{eq:MSE1}
\end{align}
in which the square of a vector, $\bs{a}$, is defined as $(\bs{a})^{2} \equiv \bs{a}^{\T} \bs{a}$.
Minimising this regularised mean squared error by demanding a vanishing gradient with respect to the weights, $\bs{w}$, yields,
\begin{align}
	\left(
	    \bs{\Phi}^{\T} \bs{\sigma}^{-2} \bs{\Phi} 
	    +
	    \bs{\lambda}^{-2}
    \right)
    \bs{w}_{\min}
	\ = \
	\bs{\Phi}^{\T} \bs{\sigma}^{-2} \bs{y},
\end{align}
in which $\bs{w}_{\min}$ is the weight vector that minimises (\ref{eq:MSE1}). The resulting (optimal) function approximation (\ref{eq:fap1}) is thus given by,
\begin{equation}
    \tilde{f} (x)
	\ = \
	\bs{y}^{\T} \bs{\sigma}^{-2} \bs{\Phi}
	\left(
	    \bs{\Phi}^{\T} \bs{\sigma}^{-2} \bs{\Phi}
	   	+
	    \bs{\lambda}^{-2}
	\right)^{-1}
	\bs{\phi}(x) .
\label{eq:approx1}
\end{equation}
The inverse is guaranteed to exist as long as the regularisation term is non-zero, i.e. $\lambda_{i}\neq0, \forall i \in \{1, \dots, N_{\text{b}}\}$.
Note that a $(N_{\text{b}} \times N_{\text{b}})$-dimensional linear system must be solved to obtain the approximate solution,
and hence the computational cost of the primal formulation is determined by the number of basis functions, $N_{\text{b}}$.

%%%%%%%%%%%%%%%%%%%%%%%%%%%%%%%%
\subsubsection{Dual formulation}
%%%%%%%%%%%%%%%%%%%%%%%%%%%%%%%%
In the dual formulation, the approximation, $\tilde{f} (x)$, is modelled as a linear combination of (evaluations of a kernel function on) the data,
\begin{align}
    \tilde{f} (x)
	\ = \
	\sum_{d=1}^{N_{\text{d}}} v_{d} \, k(x_{d}, x)
	\ \equiv \
	\bs{v}^{\T} k(\bs{x}, x),
\label{eq:fap2}
\end{align}
in which the kernel is defined in terms of the basis functions,
\begin{equation}
    k(x, x')
	\ \equiv \
	\sum_{i=1}^{N_{\text{b}}} \phi_{i}(x) \lambda_{i}^{2} \phi_{i}(x')
	\ = \
	\bs{\phi}(x)^{\T} \bs{\lambda}^{2} \bs{\phi}(x') ,
\label{eq:kernel}
\end{equation}
where we used the regularisation parameter, $\bs{\lambda}$, from equation (\ref{eq:MSE0}).
This definition of the kernel ensures the correspondence between the primal and dual formulation (see also Section \ref{subsec:pridu}).
Intuitively, the kernel expresses how the solution hinges on the data points, $\x$.
From this definition of the kernel, it can be seen that the weights of the primal and dual formulation are related as, $\bs{w} = \bs{\lambda}^{2} \bs{\Phi}^{\T} \bs{v}$. Also here, the appropriate weights can be obtained by minimising the regularised mean squared error. Keeping our previous definitions, the error (\ref{eq:MSE1}) in terms of the new weights, $\bs{v}$, reads,
\begin{align}
	\text{RMSE}(\bs{v})
	\ &\equiv \
	\left(
	    \bs{\sigma}^{-1}
	    \left(
	        \bs{\Phi} \bs{\lambda}^{2} \bs{\Phi}^{\T} \bs{v}
	        -
	        \bs{y}
	    \right)
	\right)^{2}
	+
	\left(
	    \bs{\lambda} \bs{\Phi}^{\T} \bs{v}
	\right)^{2} .
\label{eq:MSE2}
\end{align}
Minimising this regularised mean squared error by demanding a vanishing gradient with respect to the new weights, $\bs{v}$, yields,
\begin{align}
	\left(
	    \bs{\Phi} \bs{\lambda}^{2} \bs{\Phi}^{\T} \, \bs{\sigma}^{-2} \, \bs{\Phi} \bs{\lambda}^{2} \bs{\Phi}^{\T}
	    \ + \
	    \bs{\Phi} \bs{\lambda}^{2} \bs{\Phi}^{\T}
    \right)
    \bs{v}_{\min}
	\ = \
	\bs{\Phi} \bs{\lambda}^{2} \bs{\Phi}^{\T} \bs{\sigma}^{-2} \, \bs{y} ,
\label{eq:dual}
\end{align}
in which $\bs{v}_{\min}$ is the weight vector that minimises (\ref{eq:MSE2}). Note that $\bs{\Phi} \bs{\lambda}^{2} \bs{\Phi}^{\T}$ might not be invertible and thus equation (\ref{eq:dual}) might not have a unique solution.
However, we can always pick the uniquely solvable system that will also minimise (\ref{eq:MSE2}), by omitting the overall factor, $\bs{\Phi} \bs{\lambda}^{2} \bs{\Phi}^{\T}\bs{\sigma}^{-2}$, which yields,
\begin{align}
	\left(
	    \bs{\Phi} \bs{\lambda}^{2} \bs{\Phi}^{\T}
	    \ + \
	    \bs{\sigma}^{2}
    \right)
    \bs{v}_{\min}
	\ = \
	\bs{y} .
\end{align}
The resulting function approximation then reads,
\begin{equation}
    \tilde{f} (x)
	\ = \
	\bs{y}^{\T}
	\left(
	    \bs{\Phi} \bs{\lambda}^{2} \bs{\Phi}^{\T}
	    \ + \
	    \bs{\sigma}^{2}
	\right)^{-1}
	k(\bs{x}, x).
\label{eq:approx2}
\end{equation}
Note that the inverse is guaranteed to exist as long as $\sigma_{i}\neq0, \forall i \in \{1, \dots, N_{\text{d}}\}$.
In this case, a $(N_{\text{d}} \times N_{\text{d}})$-dimensional linear system needs to be solved to obtain the approximate solution,
and thus, in contrast to the primal formulation, the computational cost of the dual formulation is determined by the number of data points, $N_{\text{d}}$.

Note that the dual formulation can also be constructed directly from a given kernel without any link to a set of basis functions. In particular, the design matrix always appears as, $\bs{\Phi} \bs{\lambda}^{2} \bs{\Phi}^{\T} = k(\bs{x},\bs{x})$, and thus can always be replaced by its equivalent kernel expression.

%%%%%%%%%%%%%%%%%%%%%%%%%%%%%%%%%%%%%%%%%%%%%%
\subsubsection{Primal versus Dual formulation}
\label{subsec:pridu}
%%%%%%%%%%%%%%%%%%%%%%%%%%%%%%%%%%%%%%%%%%%%%%
One can show that the solutions of the primal (\ref{eq:approx1}) and dual (\ref{eq:approx2}) formulation are equal.
For this to be true, one needs to show that
\begin{equation}
    \bs{\sigma}^{-2} \bs{\Phi}
	\left(
	    \bs{\Phi}^{\T} \bs{\sigma}^{-2} \bs{\Phi}
	   	+
	    \bs{\lambda}^{-2}
	\right)^{-1}
	\ = \
	\left(
	    \bs{\Phi} \bs{\lambda}^{2} \bs{\Phi}^{\T}
	    +
	    \bs{\sigma}^{2}
	\right)^{-1}
	\bs{\Phi} \bs{\lambda}^{2},
\label{eq:duality}
\end{equation}
as is done in Appendix \ref{appendix:duality}. The only, yet key, difference between both formulations is thus the size of the linear system that needs to be solved to obtain the solution.

%%%%%%%%%%%%%%%%%%%%%%%%%%%%%%%%%%%%%%%%%%%%%%%%%%%%%%%%
\subsubsection{Solving linear PDEs as linear regression}
\label{subsubsec:PDE}
%%%%%%%%%%%%%%%%%%%%%%%%%%%%%%%%%%%%%%%%%%%%%%%%%%%%%%%%
Numerically solving linear operator equations, and in particular linear partial
differential equations (PDEs), can be viewed as a linear regression problem. Say we want
to numerically solve a PDE,
\begin{equation}
\begin{split}
    \mathscr{L} f(x) \ &= \ g(x), \ \ x \in D \\
    \mathscr{B} f(x) \ &= \ h(x), \ \ x \in \partial D
\end{split}
\label{eq:PDE}
\end{equation}
on a domain, $D$, with boundary, $\partial D$, where the PDE and boundary conditions are
determined respectively by the linear operators $\mathscr{L}$ and $\mathscr{B}$.
Suppose that for the numerical solution the domain is discretised to $\tilde{D}$,
and that $\bs{a}$ is a vector containing the points in $\tilde{D}$, and the
boundary is discretised to $\partial\tilde{D}$, and that $\bs{b}$ is a
vector containing the points in $\partial\tilde{D}$, then we can split the data as,
\begin{equation}
    (\bs{x}, \bs{y})
	\ = \
    \left(
    	\left(
    		\begin{matrix}
    			\bs{a} \\
    			\bs{b}
    		\end{matrix}
    	\right), \,
    	\left(
    		\begin{matrix}
				g(\bs{a}) \\
				h(\bs{b})
    		\end{matrix}
   		\right)
    \right)
	\ \equiv \
    \left(
    	\left(
    		\begin{matrix}
    			\bs{a} \\
    			\bs{b}
    		\end{matrix}
    	\right), \,
    	\left(
    		\begin{matrix}
    			\bs{g} \\
    			\bs{h}
    		\end{matrix}     
   		\right)
    \right) .
\end{equation}
Similarly, we can split the matrix, $\bs{\sigma}$, as, $\bs{\sigma} = \text{diag}(\bs{\sigma}_{\text{L}}, \bs{\sigma}_{\text{B}})$, and the design matrix, $\bs{\Phi}$, which has function evaluations at different data points on its 
rows, can be split as,
\begin{equation}
	\bs{\Phi}
	\ \equiv \
    	\left(
    		\begin{matrix}
				\mathscr{L}\bs{\phi}\left(\bs{a}\right) \\
    			\mathscr{B}\bs{\phi}\left(\bs{b}\right)
    		\end{matrix}
    	\right) 
	\ \equiv \
    	\left(
    		\begin{matrix}
    			\bs{\Phi}_{\text{L}} \\
    			\bs{\Phi}_{\text{B}}
    		\end{matrix}
    	\right) .
\label{eq:pde_design}
\end{equation}
In this new notation, the key matrices appearing in the primal and dual formulation can respectively be written as,
\begin{align}
	\bs{\Phi}^{\T} \bs{\sigma}^{-2} \bs{\Phi} + \bs{\lambda}^{-2}
	\ &= \
	\bs{\Phi}_{\text{L}}^{\T} \bs{\sigma}_{\text{L}}^{-2} \bs{\Phi}_{\text{L}}^{\phantom{\T}}
	\ + \
	\bs{\Phi}_{\text{B}}^{\T} \bs{\sigma}_{\text{B}}^{-2} \bs{\Phi}_{\text{B}}^{\phantom{\T}}
	\ + \
	\bs{\lambda}^{-2} \\
	\bs{\Phi} \bs{\lambda}^{2} \bs{\Phi}^{\T} + \bs{\sigma}^{2}
	\ &= \
   	\left(
   		\begin{matrix*}[l]
			\bs{\Phi}_{\text{L}}^{\phantom{\T}} \bs{\lambda}^{2} \bs{\Phi}_{\text{L}}^{\T} + \bs{\sigma}_{\text{L}}^{2} &
			\bs{\Phi}_{\text{L}}^{\phantom{\T}} \bs{\lambda}^{2} \bs{\Phi}_{\text{B}}^{\T}                              \\[1.1mm]
			\bs{\Phi}_{\text{B}}^{\phantom{\T}} \bs{\lambda}^{2} \bs{\Phi}_{\text{L}}^{\T}                              &
			\bs{\Phi}_{\text{B}}^{\phantom{\T}} \bs{\lambda}^{2} \bs{\Phi}_{\text{B}}^{\T} + \bs{\sigma}_{\text{L}}^{2} \\
   		\end{matrix*}
   	\right) .
   	\label{eq:pde2}
\end{align}
In terms of the kernel, equation (\ref{eq:pde2}) can also be rewritten as,
\begin{equation}
	k(\bs{x}, \bs{x}) + \bs{\sigma}^{2}
	\, = \
	\left(
   	\begin{matrix*}[l]
	    \mathscr{L}_{1} \mathscr{L}_{2} k(\bs{a}, \bs{a}) + \bs{\sigma}_{\text{L}}^{2} &
	    \mathscr{L}_{1} \mathscr{B}_{2} k(\bs{a}, \bs{b}) \\
	    \mathscr{B}_{1} \mathscr{L}_{2} k(\bs{b}, \bs{a}) &
	    \mathscr{B}_{1} \mathscr{B}_{2} k(\bs{b}, \bs{b}) + \bs{\sigma}_{\text{B}}^{2} \\
    \end{matrix*}
    \right),
\end{equation}
in which the subscripts on the operators indicate whether they act on the first or second argument of the kernel.

As with linear regression, numerically solving the PDE can thus be formulated as a minimisation problem and can be solved both in the primal and dual formulation.
The only difference is that the design matrix, $\bs{\Phi}$, should be redefined as in equation (\ref{eq:pde_design}).
This technique is known as the collocation method for solving operator equations \citep[see e.g.][]{Fasshauer1999, Schaback2006}.

Intuitively, this can be understood as follows.
The weights for the solution of the linear operator equation (\ref{eq:PDE}) are determined in terms of the basis functions, $\{\phi_{i}\}$, by fitting the functions $g(x)$ and $h(x)$, with the basis functions $\{\mathscr{L}\phi_{i}\}$ and $\{\mathscr{B}\phi_{i}\}$ respectively.
Hence, the basis functions should ideally be chosen such that $\{\mathscr{L}\phi_{i}\}$ can properly fit $g(x)$, $\{\mathscr{B}\phi_{i}\}$ can properly fit $h(x)$, and $\{\phi_{i}\}$ can properly fit the sought after solution function $f(x)$.

%%%%%%%%%%%%%%%%%%%%%%%%%%%%%%%%%%%%%%%
\subsection{Bayesian Linear Regression}
\label{subsec:BLR}
%%%%%%%%%%%%%%%%%%%%%%%%%%%%%%%%%%%%%%%
The framework of linear regression can also be derived in a Bayesian or probabilistic setting. Here we consider a stochastic function, $F(x)$, giving a probability distribution over the possible results for every value, $x$, and are interested in the distribution of this function as it is conditioned on observations of evaluations $(\bs{x},\bs{y})$ of that function, i.e. our goal is to find $p(F(x) \, | \, \bs{y})$. Note that, to simplify notation further on, we write $|\bs{y}$ to denote conditioning on the data, whereas we actually mean $|(\bs{x,\bs{y}})$.
We summarised the definitions and some further explanations of the statistical concepts that are used throughout this and subsequent sections in Appendix \ref{subsec:defs}.

%%%%%%%%%%%%%%%%%%%%%%%%%%%%%%%%%%%%%%%%%%%
\subsubsection{Bayesian primal formulation}
%%%%%%%%%%%%%%%%%%%%%%%%%%%%%%%%%%%%%%%%%%%
Given a linear model in the primal formulation with corresponding weights, $\bs{w}$, a zero-mean Gaussian error on the observed function evaluations, $\bs{\mathcal{Y}}$, results in a Gaussian likelihood given by,
\begin{equation}
    p\left( \bs{\mathcal{Y}} \, | \, \bs{w}\right)
	\ = \
	\mathcal{N}
	\left(
		\bs{w}^{\T} \bs{\phi}(\bs{x}) ,
		\,
		\bs{\sigma}^{2}
	\right)
    \ = \
	\mathcal{N}
	\left(
		\bs{\Phi} \bs{w},
		\,
		\bs{\sigma}^{2}
	\right).
\end{equation}
Note that we reused the variable $\bs{\sigma}^{2}$ and reinterpreted it as the variance on the data, allowing for deviations from the mean value, $\bs{\Phi} \bs{w}$, predicted by the model given the weights, $\bs{w}$.  We will see below that both interpretations are indeed compatible.
Furthermore, we assume a zero-mean Gaussian prior on the stochastic weights, $\bs{\mathcal{W}}$,
\begin{equation}
    p\left(\bs{\mathcal{W}}\right)
	\ = \
	\mathcal{N}
	\left(
		\bs{0}, \, \bs{\lambda}^{2}
	\right) .
\end{equation}
Note that we reused the variable $\bs{\lambda}^{2}$ and reinterpreted it as the variance of the prior on the weights.
Using the relations given in Appendix \ref{app:margcondGauss}, we can infer that the implied distribution of the weights conditioned on the data is given by,
\begin{equation}
    p(\bs{\mathcal{W}}\,|\,\bs{y})
	\ = \
	\mathcal{N}
	\left(
		\bs{\mu}_{\bs{w}|\bs{y}},
		\,
		\bs{\Sigma}_{\bs{w}|\bs{y}}
	\right),
\end{equation}
in which the mean vector and covariance matrix are defined as,
\begin{align}
    \bs{\mu}_{\bs{w}|\bs{y}}
	\ &\equiv \
	\bs{\Sigma}_{\bs{w}|\bs{y}}
		\bs{\Phi}^{\T} \bs{\sigma}^{-2} \bs{y},
	\label{eq:weights_mean} \\
    \bs{\Sigma}_{\bs{w}|\bs{y}}
	\ &\equiv \
	\left(
		\bs{\lambda}^{-2}
		+
		\bs{\Phi}^{\T} \bs{\sigma}^{-2} \bs{\Phi}
	\right)^{-1} .
	\label{eq:weights_std}
\end{align}
Since the stochastic function, $F(x)$, is a linear mapping of the weights, $F(x)=\bs{\mathcal{W}}^{\T} \bs{\phi}(x)$, the conditioned distribution reads,
\begin{equation}
    p\big(F (x) \, | \, \bs{y} \big)
    \ = \
    \mathcal{N} \left( \mu_{\text{primal}}(x), \ \sigma_{\text{primal}}^{2}(x) \right),
\end{equation}
in which the mean and variance are defined as,
\begin{align}
    \mu_{\text{primal}} (x)
	\ &\equiv \
    \bs{\mu}_{\bs{w}|\bs{y}}^{\T} \bs{\phi}(x) , \\
    \sigma_{\text{primal}}^{2} (x)
	\ &\equiv \
    \bs{\phi}(x)^{\T} \bs{\Sigma}_{\bs{w}|\bs{y}} \bs{\phi}(x) .
\end{align}
Substituting equation (\ref{eq:weights_mean}), we rediscover the primal solution (\ref{eq:approx1}) as the mean of the resulting conditioned primal distribution,
\begin{equation}
    \mu_{\text{primal}} (x)
	\ = \
	\bs{y}^{\T} \bs{\sigma}^{-2} \bs{\Phi}
	\left(
	    \bs{\Phi}^{\T} \bs{\sigma}^{-2} \bs{\Phi}
	   	+
	    \bs{\lambda}^{-2}
	\right)^{-1}
	\bs{\phi}(x) .
\label{eq:mean_primal}
\end{equation}
Furthermore, we now also have a measure for the spread in possible approximations from the variance of the conditioned distribution,
\begin{equation}
    \sigma_{\text{primal}}^{2} (x)
    \ = \
	\bs{\phi}(x)^{\T}
	\left(
	    \bs{\Phi}^{\T} \bs{\sigma}^{-2} \bs{\Phi}
	   	+
	    \bs{\lambda}^{-2}
	\right)^{-1}
	\bs{\phi}(x) .
\label{eq:std_primal}
\end{equation}
This allows us to predict an approximation for the function, $f$, based on the data, $(\bs{x}, \bs{y})$, and provide a confidence level for the result.
It should be noted that to compute the variance either for each $x$ a separate $(N_{\text{b}}\times N_{\text{b}})$-dimensional linear system needs to be solved, or that an $(N_{\text{b}}\times N_{\text{b}})$-dimensional matrix needs to be inverted explicitly.
However, since one usually does not require a high precision for an uncertainty estimate, the matrix inverse can quickly be computed in an approximate way.

%%%%%%%%%%%%%%%%%%%%%%%%%%%%%%%%%%%%%%%%%
\subsubsection{Bayesian dual formulation}
%%%%%%%%%%%%%%%%%%%%%%%%%%%%%%%%%%%%%%%%%
A similar argument can be made for the dual formulation and is typically encountered in the context of Gaussian processes \citep[see e.g.][]{Rasmussen2006}.
Since we assumed that the weights, $\bs{\mathcal{W}}$, and the errors (or spread) in the data, $\bs{\mathcal{Y}}$, both follow a zero-mean (multivariate) Gaussian distribution, the function values and the data will follow a joint (multivariate) Gaussian distribution,
\begin{equation}
    p
	\left(
    	\left[
			\begin{matrix}
    		F (x) \\
    		\bs{y}
    		\end{matrix}
		\right]
	\right)
    \ = \
    \mathcal{N}
	\left(
    	\left[
			\begin{matrix}
				    0  \\
    			\bs{0}
    		\end{matrix}
		\right]
    	, \,
		\left[
			\begin{matrix*}[l]
    			k(x, x)      & k(x, \bs{x}) \\
    			k(\bs{x}, x) & k(\bs{x}, \bs{x}) + \bs{\sigma}^{2}
    		\end{matrix*}
		\right]
    \right) ,
\label{eq:joint}
\end{equation}
where we reused the definition of the kernel (\ref{eq:kernel}).
The posterior distribution can be obtained by conditioning on the data $(\bs{x,\bs{y}})$, using the relations given in Appendix \ref{app:condGauss},
\begin{equation}
    p\big(F (x) \, | \, \bs{y} \big)
	\ = \
    \mathcal{N} \left( \mu_{\text{dual}}(x), \ \sigma_{\text{dual}}^{2}(x) \right),
\end{equation}
in which the mean and variance are respectively defined as,
\begin{align}
    \mu_{\text{dual}}(x) \ &\equiv \ k(x, \bs{x}) \left(k(\bs{x}, \bs{x}) + \bs{\sigma}^{2}\right)^{-1} \bs{y}
\label{eq:mean_dual} \\
    \sigma_{\text{dual}}^{2}(x) \ &\equiv \ k(x, x) \ - \ 
    k(x, \bs{x}) \left( k(\bs{x}, \bs{x}) + \bs{\sigma}^{2}\right)^{-1} k(\bs{x}, x) .
\label{eq:std_dual}
\end{align}
Rewriting this in terms of the design matrix, we rediscover the dual solution (\ref{eq:approx2}) as expectation of the conditioned distribution, 
\begin{equation}
    \mu_{\text{dual}} (x)
	\ = \
	\bs{\phi}(x)^{\T}
	\bs{\lambda}^{2} \bs{\Phi}^{\T}
	\Big(
		\bs{\Phi} \bs{\lambda}^{2} \bs{\Phi}^{\T}
		+
		\bs{\sigma}^{2}
	\Big)^{-1}
	\bs{y} .
\label{eq:mean_dual2}
\end{equation}
Moreover, we can similarly obtain a measure for the quality of the approximation from the variance of the conditioned distribution,
\begin{equation}
    \sigma_{\text{dual}}^{2} (x)
    = \
	\bs{\phi}(x)^{\T}
	\left(
	    \bs{\lambda}^{2}
	    - 
	    \bs{\lambda}^{2} \bs{\Phi}^{\T}
	    \Big(
		    \bs{\Phi} \bs{\lambda}^{2} \bs{\Phi}^{\T}
		    +
		    \bs{\sigma}^{2}
	    \Big)^{-1}
	    \bs{\Phi} \bs{\lambda}^{2}
	\right)
	\bs{\phi}(x) .
\label{eq:std_dual2}
\end{equation}
Again, we can predict an approximation for the function, $f$, based on the data, $(\bs{x}, \bs{y})$, and provide a confidence level for the result.
Also here, it should be noted that to compute the variance either for each $x$ a separate $(N_{\text{d}}\times N_{\text{d}})$-dimensional linear system needs to be solved, or that an $(N_{\text{d}}\times N_{\text{d}})$-dimensional matrix needs to be inverted explicitly.
However, as in the primal formulation, since one usually does not require a high precision for an uncertainty estimate, the matrix inverse can quickly be computed in an approximate way.

%%%%%%%%%%%%%%%%%%%%%%%%%%%%%%%%%%%%%%%%%%%%%%%%%%%%%%%%%%%%%%%%
\subsubsection{Bayesian primal versus Bayesian dual formulation}
%%%%%%%%%%%%%%%%%%%%%%%%%%%%%%%%%%%%%%%%%%%%%%%%%%%%%%%%%%%%%%%%
As in the non-Bayesian case, we note that in the primal formulation a $(N_{\text{b}} \times N_{\text{b}})$-dimensional linear system needs to be solved, while in the dual formulation it is a $(N_{\text{d}} \times N_{\text{d}})$-dimensional linear system.

We already showed in the non-Bayesian case (Section \ref{subsec:pridu}) that the primal and dual solutions are equal.
Using the Woodburry matrix identity, we can now also easily verify that the variances for the primal and dual Bayesian formulation are equal, since,
\begin{equation}
	\left(
		\bs{\lambda}^{-2}
		+
		\bs{\Phi}^{\T}
		\bs{\sigma}^{-2}
		\bs{\Phi}
	\right)^{-1}
	\, = \
	\bs{\lambda}^{2}
	- 
	\bs{\lambda}^{2} \bs{\Phi}^{\T}
	\Big(
		\bs{\Phi} \bs{\lambda}^{2} \bs{\Phi}^{\T}
		+
		\bs{\sigma}^{2}
	\Big)^{-1}
	\bs{\Phi} \bs{\lambda}^{2} .
\end{equation}
We can conclude that the duality also holds in the probabilistic sense, which implies for the probability distributions that,
\begin{align}
    \mathcal{N}\left(\mu_{\text{primal}} (x), \, \sigma^{2}_{\text{primal}} (x)\,\right)
    \ = \
    \mathcal{N}\left(\mu_{\text{dual}}   (x), \, \sigma^{2}_{\text{dual}}   (x)\,\right) .
\end{align}
Both formulations are thus equivalent and can therefore be used interchangeably, as long as they are both well-defined.

It should be noted that our choice of Gaussian priors was only motivated by computational convenience, and that it is not ideal.
For instance, the Gaussian distribution always assigns a non-zero probability, also to negative values of a variable.
For many physical quantities that are only positive, such as density or temperature, this is not desirable as it can lead to non-physical results.
However, this is the case for many numerical schemes, and, bearing in mind these dangers, the Gaussian distribution is a good first approximation for the uncertainties in our variables.

The Bayesian linear regression problem can alternatively also be formulated using other distributions for the priors \citep[see e.g.][for an example using Student-$t$ distributed priors]{Shah2014}, but always at the expense of computational convenience.

%%%%%%%%%%%%%%%%%%%%%%%%%%%%%%%%%%%%%%%%%%%%%%%%%%%%%%%%%%%%%%%%%%%%%%%%%%%%%%%%%%%%%%
\subsubsection{The limit of uninformative data: $\bs{\sigma} \rightarrow \bs{\infty}$}
\label{subsubsec:limit1}
%%%%%%%%%%%%%%%%%%%%%%%%%%%%%%%%%%%%%%%%%%%%%%%%%%%%%%%%%%%%%%%%%%%%%%%%%%%%%%%%%%%%%%
In order to gain more insight into these results, we consider some limiting cases.
In the limit of uninformative data, i.e. $\bs{\sigma} \rightarrow \bs{\infty}$, the uncertainty on the data is so large that conditioning on them does not change the prior distribution.
Hence, as can be seen by taking the limit, $\bs{\sigma} \rightarrow \bs{\infty}$, in equations (\ref{eq:mean_primal}, \ref{eq:std_primal}, \ref{eq:mean_dual2}, \ref{eq:std_dual2}), one finds,
\begin{align}
    \mu_{\text{primal}}(x)        &= \mu_{\text{dual}}(x) = 0, \\
    \sigma^{2}_{\text{primal}}(x) &= \sigma^{2}_{\text{dual}}(x) = 	\bs{\phi}(x)^{\T} \bs{\lambda}^{2} \bs{\phi}(x) ,
\end{align}
which is exactly the zero-mean prior distribution that we assumed.

A similar argument\footnote{The reason why a similar argument applies is the duality between the parameters $\bs{\sigma}$ and $\bs{\lambda}$. For instance, in the simplified case that $\bs{\sigma} = \sigma \mathbb{1}$ and $\bs{\lambda} = \lambda \mathbb{1}$, the parameter determining the behaviour of the model is $\sigma/\lambda$.} can be made for the limit of perfect prior knowledge, i.e. $\bs{\lambda} \rightarrow \bs{0}$, when the confidence in the prior is so large that no conditioning on any data can change it.

%%%%%%%%%%%%%%%%%%%%%%%%%%%%%%%%%%%%%%%%%%%%%%%%%%%%%%%%%%%%%%%%%%%%%%%%%%%
\subsubsection{The limit of perfect data: $\bs{\sigma} \rightarrow \bs{0}$}
\label{subsubsec:limit2}
%%%%%%%%%%%%%%%%%%%%%%%%%%%%%%%%%%%%%%%%%%%%%%%%%%%%%%%%%%%%%%%%%%%%%%%%%%%
In order to gain further insight into the results, let us ignore any effects that might be caused by uncertainties in the data and consider the limit of perfect data, i.e. $\bs{\sigma} \rightarrow \bs{0}$. The primal and dual solutions in this limit are respectively given by,
\begin{align}
    \mu_{\text{primal}}(x)
    \ &\rightarrow \
	\bs{y}^{\T} \bs{\Phi}
	\left(
	    \bs{\Phi}^{\T} \bs{\Phi}
	\right)^{-1}
	\bs{\phi}(x) , \\
    \mu_{\text{dual}}(x)
    \ &\rightarrow \
	\bs{y}^{\T}
	\left(
		\bs{\Phi} \bs{\lambda}^{2} \bs{\Phi}^{\T}
	\right)^{-1}
	\bs{\Phi} \bs{\lambda}^{2}
    \bs{\phi}(x).
\end{align}
Note that the inverses above do not necessarily exist. In particular, if $N_{\text{d}} > N_{\text{b}}$, the singular value decomposition shows that $\bs{\Phi} \bs{\lambda}^{2} \bs{\Phi}^{\T}$ must be singular and thus only the primal formulation remains, whereas if $N_{\text{d}} < N_{\text{b}}$, it follows that $\bs{\Phi}^{\T} \bs{\Phi}$ must be singular and thus only the dual formulation remains\footnote{Note, however, that the existence of the inverses of $\bs{\Phi} \bs{\lambda}^{2} \bs{\Phi}^{\T}$ and $\bs{\Phi}^{\T} \bs{\Phi}$ still depends on the choice of basis functions and the positions of the data points.}.
As a result, in the limit $\bs{\sigma} \rightarrow \bs{0}$, if $N_{\text{d}} \neq N_{\text{b}}$, the duality between the two formulations ceases to hold and only the formulation with the smallest corresponding linear system will have a unique solution.

Moreover, note that in this limit the variance of the primal formulation always vanishes. As a result, there is no probabilistic interpretation in the limit of perfect data when $N_{\text{d}} > N_{\text{b}}$ and only the primal formulation remains.
Intuitively, this can be understood since, in general, a linear regression model using $N_{\text{b}}$ basis functions cannot perfectly fit $N_{\text{d}}$ data points.
Hence, the assumption that the data can be fit perfectly, will, in general, be wrong.
The uncertainty in the data, i.e. $\bs{\sigma} \neq \bs{0}$, is required to allow for some slack in fitting the $N_{\text{d}}$ data points with only $N_{\text{b}}$ basis functions.

Similarly, in the dual formulation, the variance in the limit of perfect data, i.e. $\bs{\sigma} \rightarrow \bs{0}$, reads,
\begin{equation}
    \sigma_{\text{dual}}^{2} (x)
    \ = \
	\bs{\phi}(x)^{\T}
	\left(
	    \bs{\lambda}^{2}
	    - 
	    \bs{\lambda}^{2} \bs{\Phi}^{\T}
	    \Big(
		    \bs{\Phi} \bs{\lambda}^{2} \bs{\Phi}^{\T}
	    \Big)^{-1}
	    \bs{\Phi} \bs{\lambda}^{2}
	\right)
	\bs{\phi}(x) ,
\end{equation}
which evidently only makes sense if the inverse of $\bs{\Phi} \bs{\lambda}^{2} \bs{\Phi}^{\T}$ exists, which requires that $N_{\text{d}} \leq N_{\text{b}}$.
In the particular case that $N_{\text{d}} = N_{\text{b}}$, demanding that $\bs{\Phi} \bs{\lambda}^{2} \bs{\Phi}^{\T}$ is invertible implies that $\bs{\Phi}$ is invertible, such that also in this case the variance vanishes.
Hence, in the limit of perfect data there is only a probabilistic interpretation if $N_{\text{d}} < N_{\text{b}}$, assuming that the inverse for $\bs{\Phi} \bs{\lambda}^{2} \bs{\Phi}^{\T}$ exits.
Intuitively this can be understood from the fact that we model the spread in the distribution with the same basis functions as we use to model the function approximation. If we assume the data to be exact and if $N_{\text{d}} \geq N_{\text{b}}$, the contributions of all basis functions are fixed by the data and there are no undetermined degrees of freedom that can cause a spread in the resulting distribution conditioned on the data.

A similar argument can be made for the limit of uninformative prior knowledge, i.e. $\bs{\lambda} \rightarrow \bs{\infty}$, when the uncertainty in the prior is so large that the regression essentially fully depends on the data.

%%%%%%%%%%%%%%%%%%%%%%%%%%%%%%%%%%%%%%%
\subsection{Uncertainty Quantification}
%%%%%%%%%%%%%%%%%%%%%%%%%%%%%%%%%%%%%%%
Quantifying uncertainties is an approximate endeavour. After all, the exact solution, $f$, is required in order to determine the exact error, $\varepsilon$, that is made in a function approximation, $\tilde{f}$, since, 
\begin{equation}
    f (x)
	\ = \
	\tilde{f} (x)
	\ + \
	\varepsilon(x) . 
\label{eq:trueerror}
\end{equation}
Although it is possible to obtain highly accurate estimates for the errors in particular models \citep[see e.g.][]{Oberkampf2010}, it is crucial to note that any form of practical on-the-fly uncertainty quantification will always only be an approximation for the true error. Just as the quality of the approximation highly depends on the estimation method, so does the quality of the error.

%%%%%%%%%%%%%%%%%%%%%%%%%%%%%%%%%%%%%%%%%%%%%%%%%%%%%%%%%%%%%%%%%%%
\subsubsection{Uncertainty in the probabilistic numerical paradigm}
%%%%%%%%%%%%%%%%%%%%%%%%%%%%%%%%%%%%%%%%%%%%%%%%%%%%%%%%%%%%%%%%%%%
Following the probabilistic numerical paradigm \citep{Henning2015, Cockayne2019, Hennig2022}, we aim to quantify the uncertainty in the solution of linear operator equations by modelling the distribution over possible solutions conditioned on the data.
In particular, we will use the expectation of the conditioned distribution as our function approximation,
\begin{equation}
    \tilde{f} (x)
    \ \equiv \
    \mathbb{E}\left[F (x) \, | \, \bs{y}\right] .
\end{equation}
As a result, we can estimate the expected squared error in this approximation with the variance of the conditioned distribution,
\begin{equation}
    \tilde{\varepsilon}^{2} (x)
    \ \equiv \
    \mathbb{V}\left[F (x) \, | \, \bs{y}\right] .
\label{eq:error}
\end{equation}
This can be inferred from the fact that the stochastic function, $F(x)$, with corresponding stochastic error, $\mathcal{E}(x)$, ought to be related as,
\begin{equation}
    F (x)
	\ = \
	\tilde{f} (x)
	\ + \
	\mathcal{E}(x) ,
\end{equation}
and the definition of the variance, which implies that,
\begin{equation}
    \mathbb{V}\left[F (x) \, | \, \bs{y}\right]
    \ = \
    \mathbb{E} \left[ \left(F(x) - \tilde{f} (x)\right)^{2} \, | \, \bs{y} \right]
    \ = \
    \mathbb{E} \left[ \mathcal{E}(x)^{2} \, | \, \bs{y} \right] .
\end{equation}
Assuming that the probabilistic model, $F (x) \, | \, \bs{y}$, is an adequate model for the actual function, $f(x)$, the variance thus quantifies the expected squared error in the function approximation.
Note that in our particular case, where the posterior is a Gaussian, the variance does not depend on the function values, $\bs{y}$, of the data but only on the locations at which the function was evaluated, $\bs{x}$.

Based on the variance in the dual formulation (\ref{eq:std_dual}), one can derive an upper and lower bound on the expected squared error,
\begin{equation}
    0 \ \leq \ \tilde{\varepsilon}^{2} (x) \ \leq \ k(x,x),
\end{equation}
where, in the left inequality, we used that the variance has to be positive and in the right inequality that $k(\bs{x}, \bs{x}) + \bs{\sigma}^{2}$ is a positive definite matrix, such that the second term in (\ref{eq:std_dual}) is always negative.
It might seem odd to have an error measure that is bounded from above.
However, one should note that it is not an upper bound on the actual error, but rather an upper bound on the expected error.

There is an alternative way to understand the error measure defined in (\ref{eq:error}) using the reproducing kernel Hilbert space (RKHS) of the kernel \citep[see e.g.][]{Cockayne2017}.
Let $\mathcal{H}$ denote the RKHS of the kernel defined in (\ref{eq:kernel}), with an associated inner product, $\langle \rangle_{\mathcal{H}}$, and norm $\|\cdot\|_{\mathcal{H}} \equiv \sqrt{\langle \cdot, \cdot\rangle_{\mathcal{H}}}$.
If we now consider the projection $Pf\in\mathcal{H}$ of the function, $f$, in the RKHS, $\mathcal{H}$, one can derive the following bound (see Appendix \ref{app:RKHSbound}),
\begin{equation}
    \left|Pf(x) - \tilde{f}(x)\right| \ \leq \ \left\|Pf\right\|_{\mathcal{H}} \ \tilde{\varepsilon}(x) .
\label{eq:RKHSbound}
\end{equation}
This means that $\tilde{\varepsilon}(x)$ bounds the local error in the approximation, $\tilde{f}$, as measured in the RKHS.
To intuitively see how this comes about without needing the notion of a RKHS, note that if the function that one tries to approximate is $f(x) = k(\x, x)$, based on the data $\y = k(\x,\x)$, then the variance (\ref{eq:std_dual}) is exactly equal to the difference between $f$, and its approximation (\ref{eq:mean_dual}).
Now if the function one tries to approximate is not $k(\x, x)$, but a linear combination of evaluations of $k(\x,x)$, this difference can grow by an additional factor which can be bounded by $\left\|Pf\right\|_{\mathcal{H}}$, yielding the bound (\ref{eq:RKHSbound}).

It should be emphasised that inequality (\ref{eq:RKHSbound}) only bounds the error in the approximation with respect to the projection of the true solution in the RKHS, $Pf$, and not the error with respect to the true solution, $f$, itself.
Hence, the strength of this bound crucially depends on the RKHS, and thus on the particular kernel, or equivalently, on the particular set of basis functions that is used.
If the projection, $Pf$, in the RKHS is a good approximation for the true function, $f$, then the error bound (\ref{eq:RKHSbound}) can also be used to bound the true error in equation (\ref{eq:trueerror}).
However, this assumes a certain regularity of the function, $f$, and again crucially depends on the particular choice of kernel or basis functions. 

If $\{\phi_{i}\}$ is a finite and orthonormal set of square-integrable basis functions on some domain, $D$, i.e. $\langle \phi_{i}, \phi_{j} \rangle \equiv \int_{D} \phi_{i} \phi_{j} = \delta_{ij}$, then the function space spanned by these basis functions is a RKHS, say $\mathcal{H}$, with reproducing kernel (\ref{eq:kernel}), with respect to the following inner product.
Since every function in the RKHS can be expressed as a linear combination of the basis functions, the inner product between $g(x)=\bs{g}^{\T} \bs{\phi}(x) \in  \mathcal{H}$ and $h(x)=\bs{h}^{\T} \bs{\phi}(x) \in  \mathcal{H}$ can be defined as $\langle g, h \rangle_{\mathcal{H}} \equiv \bs{g}^{\T} \bs{\lambda}^{-2} \bs{h}$. Hence, for a finite set of orthonormal basis functions, it is this inner product that must be used to compute the norm in the local error bound (\ref{eq:RKHSbound}).

%%%%%%%%%%%%%%%%%%%%%%%%%%%%%%%%%%%%%%%%%%%%%%%
\subsection{Example basis functions \& kernels}
\label{subsec:examples}
%%%%%%%%%%%%%%%%%%%%%%%%%%%%%%%%%%%%%%%%%%%%%%%
Given the data, the key parameters that determine the regression model are the set of basis functions or the kernel in the primal and dual formulation respectively.
In order to gain more insight into the linear regression method, we consider some specific examples of basis functions and their corresponding kernels.

%%%%%%%%%%%%%%%%%%%%%%%%%%%%%
\subsubsection{Fourier basis}
%%%%%%%%%%%%%%%%%%%%%%%%%%%%%
As a first example, consider the set of $N_{\text{b}}=2N+1$ real Fourier basis functions, $\{1\} \cup \{\sin(\omega_{n} x)\}_{n=1}^{N} \cup \{\cos(\omega_{n} x)\}_{n=1}^{N}$, where we defined $\omega_{n} \equiv 2\pi n/L$, with $L$ the size of the domain that we are interested in.
Given these basis functions, the primal representation of the function approximation (\ref{eq:approx1}) corresponds to the (truncated) Fourier series of the function that we are looking for.
If we denote the entry in $\bs{\lambda}$ corresponding to the constant with $\lambda$, the entries corresponding to the sines with $\lambda_{n}$, and the entries corresponding to the cosines with $\lambda'_{n}$, the resulting kernel can be expanded as,
\begin{equation}
\begin{split}
    k(x,x')
    \ = \
    \lambda^{2}
    \ &+ \
    \sum_{n=1}^{N}
    \left(
    \frac{\lambda_{n}^{2}+\lambda_{n}^{' 2}}{2}
    \right)
    \cos\big(\omega_{n} (x-x') \big) \\
    \ &+ \
    \sum_{n=1}^{N}
    \left(
    \frac{\lambda_{n}^{2}-\lambda_{n}^{' 2}}{2}
    \right)
    \cos\big(\omega_{n} (x+x') \big) .
\end{split}
\label{eq:Fourierkernel}
\end{equation}
Typically, one would expect that modes corresponding to the same length scale, i.e. same $n$, would have similar weights, $\lambda_{n} \approx \lambda_{n}'$, such that the second summation vanishes. This approximately renders the kernel into a radial basis function $k(x,x') \approx K(\|x-x'\|)$.

Now consider the simplest case, when all $\lambda_{n}=\lambda_{n}'=\lambda$. The kernel is then a radial basis function and can be computed explicitly,
\begin{align}
    k(x,x')
    \ &= \
    \frac{\lambda^{2}}{2}
    \left(
        1 + \frac{\sin\big(\pi(2N+1)(x-x')/L\big)}{\sin\big(\pi(x-x')/L\big)}
    \right) ,
\label{eq:Fourier_kernel}
\end{align}
which is commonly known as (one half plus) the Dirichlet kernel. The kernel attains its maximum on the diagonal, $k(x,x)=\lambda^{2}(N+1)$, and oscillates and decays away from there. The dual solution (\ref{eq:approx2}) is a linear combination of these radial basis functions centred, and thus peaking, around the data points, and decaying away elsewhere.

Note that, as the number of basis functions increases, $N \rightarrow \infty$, the kernel becomes more narrow and peaked, and in the limit tends towards a delta distribution.
Since the kernel centred around a data point represents the influence of that data point on the solution, this implies that with increasing $N$, the effect of each individual data point on the solution decreases and becomes ever more confined to a shrinking region around each data point. Similarly, with increasing $N$, the variance (or corresponding uncertainty estimate) in between data points will increase.
Intuitively, this can be understood, since increasing the number of basis functions, while keeping the number of data points fixed, will imply that the basis functions are ever less constrained by the data, a problem commonly known as over-fitting.

Over-fitting can be cured with regularisation by damping the higher order modes  in the kernel through $\bs{\lambda}$, making it less peaked in the limit of large $N$.
Note that the entries of the regularisation vector, $\bs{\lambda}$, appear as the Fourier coefficients of the kernel (\ref{eq:Fourierkernel}).
This illustrates the crucial interplay between the choice of basis functions and regularisation.
In a sense, regularisation effectively comes down to re-scaling the basis functions, since by making the re-scaling, $\phi_{i} \rightarrow \lambda_{i}\phi_{i}$, the regularisation vector can always be cast into the trivial form $\bs{\lambda}= \mathbb{1}$.

The Fourier basis allows us to restrict the desired solution of the regression problem to a minimal length scale, defined by $L/N$.
Therefore, by considering only a limited number of basis functions, $N_{\text{b}} \ll N_{\text{d}}$, one can obtain a large-scale ($L/N$) approximation for the model, which can efficiently be solved in the primal formulation, as it only requires the solution of an ($N_{\text{b}} \times N_{\text{b}}$)-dimensional system.

A potential problem with the Fourier basis is that the effective width, $\xi$, of the kernel (\ref{eq:Fourierkernel}), which can be estimated as, $\xi \sim L/N$, is the same around every data point. This is fine as long as the distance between the data points, $\x$, is much smaller than the effective width of the kernel, but causes problems, for instance, if there are ``lonely'' data points whose nearest neighbours are much farther away than the effective width of the kernel. Since the kernel rapidly decays for length scales beyond its effective width, the solution around a so-called ``lonely'' data point, say $x_{l}$, can be approximated as,
\begin{align}
    \tilde{f} (x)
	\ = \
	\bs{v}^{\T} k(\bs{x}, x)
	\ \approx \
	v_{l} k(x_{l}, x),
\end{align}
which holds, as long as $x$ is much closer to $x_{l}$ than to its nearest neighbour, say $x_{l}^{\text{n}}$, i.e. $\|x-x_{l}\| \ll \|x-x_{l}^{\text{n}}\|$.
Since the nearest neighbour is much farther away than the effective width of the kernel (by definition of a ``lonely point''), there is a significant region around $x_{l}$, defined by $\{x : \xi < \|x-x_{l}\| \ll \|x-x_{l}^{\text{n}}\| \}$, for which, 
\begin{align}
    \tilde{f} (x)
	\ \approx \
	0 .
\end{align}
Similarly, the variance (or uncertainty estimate) for the function approximation in that region will attain its maximum value,
\begin{align}
    \tilde{\varepsilon}^{2} (x)
	\ \approx \
	k(x,x)
	\ = \
	\lambda^{2}(N+1) .
\end{align}
Hence, the approximation is probably not good in that region.
This type of problem can be avoided by choosing basis functions or a kernel that is locally adapted to the distribution of data points.

%%%%%%%%%%%%%%%%%%%%%%%%%%%%%%%%%%%%%%
\subsubsection{Radial basis functions}
%%%%%%%%%%%%%%%%%%%%%%%%%%%%%%%%%%%%%%
As a second example, consider a set of basis functions generated by a radial basis function, $\psi$, centred around each data point, $x_{i}$, such that there is one basis function, $\phi_{i}$ for each data point $x_{i}$, with,
\begin{align}
    \phi_{i}(x) \ = \ \psi\left(\frac{\|x-x_{i}\|}{\xi_{i}}\right) ,
\label{eq:RBF}
\end{align}
in which $\xi_{i}$ controls the effective width of the radial basis function around data point, $x_{i}$.
The corresponding kernel for this basis reads,
\begin{align}
    k(x, x')
    \ &= \
    \sum_{i=1}^{N_{\text{b}}}
    \psi\left(\frac{\|x-x_{i}\| }{\xi_{i}}\right)
    \lambda_{i}^{2} \,
    \psi\left(\frac{\|x'-x_{i}\|}{\xi_{i}}\right) .
\end{align}
Issues with ``lonely'' data points as encountered with the Fourier basis can be avoided here by tailoring the basis functions to the data, for instance, by choosing $\xi_{i} = \|x_{i} - x_{i}^{\text{n}} \|$, in which $x_{i}^{\text{n}}$ is the nearest neighbour of $x_{i}$.

Radial basis functions are a popular choice for solving operator equations. Although, often, some care is required to cope with the ill-conditioning of the resulting linear system  \citep[see e.g.][]{Fronberg2015}.
Moreover, radial basis functions often offer an intuitive interpretation. For instance, when dealing with smoothed-particle hydrodynamics data \citep{Gingold1977, Lucy1977}, the basis functions can be related to the smoothing kernels and represent the proliferation of the data from each particle.

Since, in this approach, the basis functions are tied to the data points, approximating the solution around certain data points can be achieved by discarding the corresponding basis functions. In \cite{DeCeuster2020}, a mesh reduction method for radiative transfer models was proposed in which, based on a certain heuristic, data points which  were not deemed essential were discarded from the model.
A similar but improved reduction scheme can be obtained using the linear regression approach with basis functions tied to the data, by discarding the corresponding basis functions instead.
In this way, the data itself does not have to be discarded and can still be taken into account, while the model is reduced in computational complexity.
Furthermore, the Bayesian linear regression method can provide an estimate for the uncertainty on the result after solving the reduced model.
This leads us to the question whether there are even better bases to compress radiative transfer models.

%%%%%%%%%%%%%%%%%%%%%%%%%%%%%
\subsubsection{Wavelet bases}
%%%%%%%%%%%%%%%%%%%%%%%%%%%%%
Wavelets have a proven track record for data compression in various applications, such as sound and image processing \citep[see e.g.][]{Vetterli2001}, and have already successfully been applied to solve operator equations \citep[see e.g.][and the references therein]{Stevenson2009}.
They combine the localisation in scale of the Fourier basis, i.e. certain basis functions describe certain length scales, with the localisation in space of radial basis functions, i.e. certain basis functions describe certain regions in space. As a result, wavelet bases are  of the form,
\begin{equation}
    \psi_{m n}(x) \ = \ a^{-m/2} \ \psi \left( \frac{x - n b}{a^{m}} \right),
\end{equation}
indexed by two indices, in which $m$ describes the length scale, and $n$ describes the location, parametrised by the constants $a$ and $b$ respectively.
By imposing the mathematical structure of a multi-resolution analysis, relations between the different scales can be derived which allow one to construct orthogonal wavelet bases, which give rise to efficient algorithms to decompose functions into their wavelet components \citep[see e.g.][]{Daubechies1992}.

By selecting (or disregarding) certain wavelet basis functions we thus can locally refine (or coarsen) the solution of the linear regression problem. However, this assumes that we know where we want to refine the model and where a coarser representation suffices.
Alternatively, by expressing the data directly in a wavelet basis (e.g. using a fast wavelet transform), one can select only those components that significantly contribute \citep[see e.g.][]{Daubechies1992}.

There is a large variety of wavelet bases and the key remains to choose an appropriate one.
Moreover, when the aim is to solve operator equations, the wavelet basis should still be appropriate when acted upon with the relevant operator. Choosing appropriate wavelet bases adapted to a particular operator turns out to be a challenging endeavour \citep[see e.g.][]{Stevenson2009}.
However, recently, significant progress has been made, for instance, by \cite{Owhadi2017}, which makes wavelet bases an attractive choice for solving large (Bayesian) linear regression problems (see also Section \ref{sec:discussion}).

%%%%%%%%%%%%%%%%%%%%%%%%%%%%%%%%%%%%%%%%%%
\section{Bayesian Radiative Transfer}
\label{sec:prob_rad_trans}
%%%%%%%%%%%%%%%%%%%%%%%%%%%%%%%%%%%%%%%%%%
We can now apply the probabilistic numerical approach developed above to the particular case of radiative transfer problems.
The goal is to find the radiation field throughout a region, based on the radiative properties of the medium and some boundary conditions. 
The radiation field can be described by the specific monochromatic intensity, $I_{\nu}(\x, \n)$, i.e. the energy at a point, $\x$, transported in a direction, $\n$, in a certain frequency bin, $\nu$.
Interactions between the radiation field and the medium can be described in terms of the change they imply in the specific monochromatic intensity.
The radiative transfer equation is a linear operator equation that relates this change to the radiative properties of the medium,
\begin{equation}
    \mathscr{L} I_{\nu}(\x, \n) \ = \ \eta_{\nu}(\x) ,
\label{eq:RTE}
\end{equation}
in which, $\eta_{\nu}(\x)$, is the emissivity of the medium.
In the time-independent case and including scattering, the operator, $\mathscr{L}$, acts on the intensity as \citep[see e.g.][]{Mihalas1984},
\begin{equation}
\begin{split}
    \mathscr{L} I_{\nu}(\x,\n)
    \ \equiv \
    &\Big(\chi_{\nu}(\x) \ + \ \n \cdot \nabla \Big) \, I_{\nu}(\x,\n) \\
    & \ - \ \oint \D \Omega' \int_{0}^{\infty} \D \nu' \ \Phi_{\nu \nu'} \left(\x,\n, \n'\right) I_{\nu'}(\x,\n') .
\end{split}
\label{eq:op_L}
\end{equation}
Here, we introduced the opacity, $\chi_{\nu}(\x)$, and the scattering redistribution function, $\Phi_{\nu \nu'} \left(\x, \n, \n'\right)$.
Since $\mathscr{L}$ is a linear operator, the solution of the radiative transfer equation, given appropriate boundary conditions, can be viewed as a Bayesian linear regression problem.
It remains to find an appropriate set of basis functions (in the primal formulation), or to find an appropriate kernel (in the dual formulation), given the radiative properties of the medium.

%%%%%%%%%%%%%%%%%%%%%%%%%%%%%%%%%%%%%%%%%%%%%%%%%%
\subsection{Approximate Radiative Transfer models}
%%%%%%%%%%%%%%%%%%%%%%%%%%%%%%%%%%%%%%%%%%%%%%%%%%
Almost all astrophysical simulations require some kind of radiative transfer model.
However, due to the significant computational cost, one is often forced to make drastic approximations.
In this section, we show how the primal formulation can be used to create reduced-order or approximate radiative transfer models and show how it can be applied, for instance, to compute approximated Lambda operators for atomic and molecular line transfer.

%%%%%%%%%%%%%%%%%%%%%%%%%%%%%%%%%%%%
\subsubsection{Reduced-order models}
\label{subsubsec:rom}
%%%%%%%%%%%%%%%%%%%%%%%%%%%%%%%%%%%%
As already alluded to in Section \ref{subsec:examples}, we can obtain approximate solutions for a linear regression problem by considering reduced sets of basis functions in the primal formulation.
The basis functions essentially map the regression problem to an $N_{\text{b}}$-dimensional feature space in which the problem is solved.
Therefore, in a sense, the primal solution (\ref{eq:approx1}) can be interpreted as follows,
\begin{equation}
    \tilde{f} (x)
	\ = \
	\bs{y}^{\T}
	{\underbrace{
	\vphantom{\left(\bs{\Phi}^{\T} \bs{\sigma}^{-2} \bs{\Phi} +\bs{\lambda}^{-2}\right)^{-1}}
	\bs{\sigma}^{-2} \bs{\Phi}}_{\text{compress}}
	}
	{\underbrace{
	\left(
	    \bs{\Phi}^{\T} \bs{\sigma}^{-2} \bs{\Phi}
	   	+
	    \bs{\lambda}^{-2}
	\right)^{-1}
	}_{\text{solve}}
	}
	{\underbrace{
	\vphantom{\left(\bs{\Phi}^{\T} \bs{\sigma}^{-2} \bs{\Phi} +\bs{\lambda}^{-2}\right)^{-1}}
	\bs{\phi}(x)}_{\text{decompress}}
	} .
\label{eq:comp}
\end{equation}
First, the $N_{\text{d}}$-dimensional data vector, $\bs{y}$, is mapped into the $N_{\text{b}}$-dimensional feature space, which can be viewed as a projection or compression, if $N_{\text{b}} < N_{\text{d}}$.
Then, the problem is solved in the $N_{\text{b}}$-dimensional feature space, and finally mapped back into the desired format.
The least-squares problem posed in equation (\ref{eq:MSE1}) minimises the compression loss.
The resulting reduced-order model provides an approximate solution to the (more) exact radiative transfer problem, in contrast to the exact solutions to approximate models that are often used.
Moreover, the probabilistic interpretation allows us to quantify with the variance (\ref{eq:std_primal}) the uncertainty that was introduced by compressing the model, allowing us to strictly control the trade-off between accuracy and computational cost.

By denoting the first part of equation (\ref{eq:comp}) as a compression of the data, one could ask whether the vector-matrix multiplication in equation (\ref{eq:comp}) is the most efficient way to perform this compression.
Indeed, for the Fourier and wavelet bases there exist more efficient algorithms to express a given data set into these bases, the so-called Fast Fourier Transform (FFT) and Fast Wavelet Transform (FWT) respectively \citep[see e.g.][]{Press2007}.
These can reduce the computational cost of these models even further.

The type and amount of compression critically depends on the set of basis functions that is used.
One way to choose them, for instance, is by performing a principle component analysis on the design matrix, $\bs{\Phi}$.
This yields what is known as a proper orthogonal decomposition \citep[POD; see e.g.][]{Benner2017}.
In addition to performing this compression, the probabilistic approach now also allows to quantify the uncertainties that are thus introduced.

%%%%%%%%%%%%%%%%%%%%%%%%%%%%%%%%%%%%%%%%%%%%
\subsubsection{Application: Approximate Lambda Operators}
%%%%%%%%%%%%%%%%%%%%%%%%%%%%%%%%%%%%%%%%%%%%
Approximations to radiative transfer are often used to accelerate iterative line radiative transfer solvers.
Models involving atomic or molecular line radiative transfer show a non-linear coupling between the radiative properties of the medium and the radiation field.
This coupling can be expressed as, 
\begin{equation}
    I \ = \ \Lambda \left[\eta\left(I\right)\right],
\label{eq:Lambda}
\end{equation}
in which $I$ indicates the radiation field, $\Lambda$ is a linear operator, and we explicitly indicated the dependence of $\eta$ on the radiation field. It is this dependency of $\eta$ on $I$ that is usually non-linear.
Due to this non-linear coupling, the radiation field has to be computed in an iterative way, \citep[see e.g. Chapter 13 in][]{Hubeny2014},
\begin{equation}
    I^{(n+1)} \ = \ \Lambda \left[\eta\left(I^{(n)}\right)\right] .
\end{equation}
This iterative scheme often shows notoriously slow convergence and one often has to resort to acceleration techniques, such as operator splitting \citep{Cannon1973_angle, Cannon1973_frequency}, which yields the implicit scheme,
\begin{equation}
    I^{(n+1)} \ = \
    \Lambda^{*} \left[\eta\left(I^{(n+1)}\right)\right]
    \ + \
    \left( \Lambda - \Lambda^{*} \right) \left[\eta\left(I^{(n)}\right)\right] ,
\label{eq:op-split}
\end{equation}
in which the linear operator, $\Lambda^{*}$, is an approximation for the operator, $\Lambda$, that can easily be inverted \citep[see e.g.][for a specific implementation]{Rybicki1991}.
Intuitively, the better the approximation, $\Lambda^{*}$, the smaller the dependence on the previous iteration in (\ref{eq:op-split}), and thus the better convergence will be.
The key to success in this acceleration scheme is to find a good approximate operator, $\Lambda^{*}$.

Comparing equations (\ref{eq:approx1}), (\ref{eq:RTE}), and (\ref{eq:Lambda}), one can see that, in the primal formulation, the operator, $\Lambda$, in matrix form is given by,
\begin{align}
    \bs{\Lambda}
    \ &= \
    \bs{\phi}(x)^{\T}
	\bs{\lambda}^{2} \bs{\Phi}^{\T}
	\Big(
		\bs{\Phi} \bs{\lambda}^{2} \bs{\Phi}^{\T}
		+
		\bs{\sigma}^{2}
	\Big)^{-1} .
\end{align}
As a result, good approximations to this operator can be obtained, for instance, by considering reduced sets of basis functions in the corresponding linear regression problem, as shown in Section \ref{subsubsec:rom}.

%%%%%%%%%%%%%%%%%%%%%%%%%%%%%%%%%%%%%%
\subsection{Method of Characteristics}
\label{subsec:characteristics}
%%%%%%%%%%%%%%%%%%%%%%%%%%%%%%%%%%%%%%
In order to make the probabilistic approach to radiative transfer more concrete, we consider the specific example of the method of characteristics and derive it from a Bayesian point of view.

In its simplest form, in the absence of scattering and neglecting any frequency dependence, the time-independent radiative transfer equation along a single ray reads,
\begin{equation}
	\mathscr{L}_{\change{s}} I(s)
	\ = \
	\eta(s) ,
\end{equation}
in which the linear differential operator, $\mathscr{L}_{\change{s}}$, is defined as,
\begin{equation}
	\mathscr{L}_{\change{s}}
	\ \equiv \
	\chi(s) \ + \ \partial_{s} .
\label{eq:RTE_DO}
\end{equation}
For future reference, we already note that the Green's function for this linear operator, $\mathscr{L}_{\change{s}}$, is given by,
\begin{equation}
	G(z, s)
	\ = \
	\Theta(s-z) \ e^{-\tau(z, \, s)},
\label{eq:Green}
\end{equation}
in which $\Theta$ is the Heaviside function, and the optical depth, $\tau$, over an interval $[z,s]$ along the ray, is defined as,
\begin{equation}
	\tau(z, \, s) \ \equiv \ \int_{z}^{s} \D s' \ \chi(s') ,
\label{eq:optical_depth}
\end{equation}
such that $\partial_{s}\tau(z, \, s) = \chi(s)$, and thus, as expected,
\begin{align}
	\mathscr{L}_{s} G(z, \, s)
	\ = \
	\delta(s - z) .
\label{eq:def_Green}
\end{align}
Using this Green's function one can (at least formally) solve the radiative transfer equation, as in the method of characteristics.

%%%%%%%%%%%%%%%%%%%%%%%%%%%%%%%%%%%%%%%%%%%%%%%%%%%
\subsubsection{Classical method of characteristics}
%%%%%%%%%%%%%%%%%%%%%%%%%%%%%%%%%%%%%%%%%%%%%%%%%%%
The method of characteristics solves the transfer equation starting from its formal solution based on the Green's function. Given the boundary condition, $I(s_{0})=I_{0}$, at boundary point, $s_{0}$, one finds,
\begin{align}
	I(s)
	\ = \
	I_{0} \, e^{-\tau(s_{0}, s)}
	\ + \
	\int_{s_{0}}^{s} \D s' \
	\eta(s') \, e^{-\tau(s',s)} .
\label{eq:formal_solution}
\end{align}
The required integrals in equations (\ref{eq:optical_depth}) and (\ref{eq:formal_solution}) are then evaluated using a (local) interpolation both for the emissivity and opacity functions, $\eta(s)$ and $\chi(s)$.

At this point, a distinction is often made between so-called short and long characteristic methods depending on the location of the point $s_{0}$ in the discretisation.
In the case of short characteristics, $s_{0}$ is taken to be the previous point in the discretisation, while, in the case of long characteristics, it is taken to be the boundary of the computational domain.
For our intents and purposes this distinction does not matter, so we continue with the formulation in (\ref{eq:formal_solution}), in which $s_{0}$ can be any point in the discretisation.

The emissivity and opacity are usually interpolated using a linear scheme. Given a kernel, $\kappa$, the interpolant (\ref{eq:approx2}) in the dual formulation can be written as,
\begin{align}
	\tilde{\eta}(s)
	\ &\equiv \
	\bs{\eta}^{\T} \left(\kappa(\bs{a}, \bs{a}) + \bs{\sigma}_{\eta}^{2}\right)^{-1} \kappa(\bs{a}, s) ,
\label{eq:int_emi} \\
	\tilde{\chi}(s)
	\ &\equiv \
	\bs{\chi}^{\T} \left(\kappa(\bs{a}, \bs{a}) + \bs{\sigma}_{\chi}^{2}\right)^{-1} \kappa(\bs{a}, s) ,
\label{eq:int_opac}
\end{align}
in which $\bs{a}$ is the vector of positions at which the values for $\eta$ and $\chi$ are given, and where $\bs{\sigma}^{2}_{\eta}$ and $\bs{\sigma}^{2}_{\chi}$ denote the diagonal matrices with the variances for the given values of $\eta$ and $\chi$ respectively.
However, in the classical method of characteristics, these variances are never used, and thus implicitly assumed to be negligible.
In the Bayesian method of characteristics, however, they model the uncertainties in the emissivities and opacities that originate, for instance, from the uncertainties in the radiative data (see Section \ref{subsec:bayesian-moc}).
Furthermore, we note that, in principle, one can use a different kernel for $\eta$ and $\chi$, although, in practice, one often uses the same one.
One particularly popular choice of kernel is the one corresponding to the basis of Lagrange polynomials, since they trivially satisfy the interpolation property.
With equations (\ref{eq:int_emi}) and (\ref{eq:int_opac}), the formal solution yields,
\begin{equation}
	\tilde{I} (s)
    \ = \
	I_{0} \, e^{-\tilde{\tau}(s_{0},s)}
	\ + \
	\bs{\eta}^{\T} \, \bm{K}_{\eta}^{-1}
    \int_{s_{0}}^{s} \D s' \
    \kappa(\bs{a}, s') \
    e^{-\tilde{\tau}(s',s)}
\label{eq:characteristics}
\end{equation}
in which the interpolated optical depth is given by,
\begin{align}
	\tilde{\tau}(z, \, s)
	\ &= \
	\bs{\chi}^{\T} \bm{K}_{\chi}^{-1} \int_{z}^{s} \D s' \ \kappa(\bs{a}, s') ,
\label{eq:optdepth}
\end{align}
where, for brevity, we defined the matrices $\bm{K}_{\eta} \equiv \kappa(\bs{a}, \bs{a}) + \bs{\sigma}_{\eta}^{2}$, and $\bm{K}_{\chi} \equiv \kappa(\bs{a}, \bs{a}) + \bs{\sigma}_{\chi}^{2}$.
The integrals in equations (\ref{eq:characteristics}) and (\ref{eq:optdepth}) can now be evaluated on the (analytically) known kernel function, $\kappa$, thus solving the radiative transfer equation.

%%%%%%%%%%%%%%%%%%%%%%%%%%%%%%%%%%%%%%%%%%%%%%%%%%
\subsubsection{Bayesian method of characteristics}
\label{subsec:bayesian-moc}
%%%%%%%%%%%%%%%%%%%%%%%%%%%%%%%%%%%%%%%%%%%%%%%%%%
Now we show how the method of characteristics can be derived as a Bayesian linear regression problem in the dual formulation by choosing a particular type of kernel, or equivalently by choosing a particular set of basis functions.

Given the Green's function (\ref{eq:Green}) for the differential operator in the radiative transfer equation, consider a kernel of the form,
\begin{align}
	k(z, s)
	\ &= \
	\int_{-\infty}^{+\infty} \D s'
	\int_{-\infty}^{+\infty} \D z' \
	\kappa(s', z') \
	G(z', z) \
	G(s', s) ,
\label{eq:kernelGreen}
\end{align}
in which $\kappa(s', z')$ is another kernel from which we only demand that it does not correlate the region $s>s_{0}$ with $s<s_{0}$. The reason for this is, that, in the classical method of characteristics, we want to use the solution at $s_{0}$ as a true boundary condition, i.e. such that nothing at $s>s_{0}$ affects the solution at $s<s_{0}$, and vice versa.
This implies a block diagonal kernel of the from,
\begin{equation}
\begin{split}
	\kappa(z,s)
	\ &\equiv \
	\Theta(s_{0}-z) \Theta(s_{0}-s) 
    \kappa(z,s)  \\
	& \quad \ + \
    \Theta(z-s_{0}) \Theta(s-s_{0})
    \kappa(z,s) .
\end{split}
\label{eq:splitcond}
\end{equation}
If we now assume that $\forall a \in \bs{a}: a > s_{0}$, and we assume no error on the boundary condition, one can show (see Appendix \ref{appendix:Green}) that the dual solution for the Bayesian linear regression problem reads, 
\begin{equation}
	\tilde{I} (s)
    \ = \
	I_{0} \, e^{-\tilde{\tau}(s_{0},s)}
	\ + \
	\bs{\eta}^{\T} \, \bm{K}_{\eta}^{-1}
    \int_{s_{0}}^{s} \D s' \
    \kappa(\bs{a}, s') \
    e^{-\tilde{\tau}(s',s)}
\label{eq:characteristics_prob}
\end{equation}
with the corresponding uncertainty estimate given by,
\begin{equation}
\begin{split}
    \tilde{\varepsilon}_{I}^{2} (s)
    \ &= \
	\int_{s_{0}}^{s} \D s'
	\int_{s_{0}}^{s} \D z' \
	e^{-\tilde{\tau}(s', \, s)}
	e^{-\tilde{\tau}(z', \, s)} \\
	& \ \ \ \ \ \times
	\left(
	\kappa(s',\,z')
	-
	\kappa(\bs{a},\,s')^{\T}
	\bm{K}_{\eta}^{-1}
    \kappa(\bs{a},\,z')
	\right) .
\end{split}
\label{eq:characteristics_std}
\end{equation}
Note that the probabilistic solution (\ref{eq:characteristics_prob}) is exactly the same as the classical solution (\ref{eq:characteristics}) for the method of characteristics.
Therefore, we can conclude that both methods are equivalent, but with the important difference that the probabilistic approach can account for uncertainties on the input (through $\bs{\sigma}_{\eta}$) and we thus can estimate the uncertainty on the result.
Moreover, in the expression between parentheses in equation (\ref{eq:characteristics_std}),
\begin{equation}
	\kappa(s',\,z')
	-
	\kappa(\bs{a},\,s')^{\T}
	\bm{K}_{\eta}^{-1}
    \kappa(\bs{a},\,z')
\end{equation}
we recognise the resulting variance in the dual formulation (\ref{eq:std_dual}) that stems from the interpolation of the emissivity (\ref{eq:int_emi}).

We should note that in the definition of the kernel (\ref{eq:kernelGreen}), we implicitly assumed that we knew the Green's function (\ref{eq:Green}), and thus we implicitly assumed that we knew the optical depth (\ref{eq:optical_depth}).
In general, we do not have an exact expression for the optical depth.
However, we can find an approximate solution by solving another Bayesian linear regression problem for the operator equation,
\begin{align}
    \partial_{s}\tau(z, \, s) \ = \ \chi(s) ,
\end{align}
with boundary condition, $\tau(z, \, z) = 0$, which, using the kernel,
\begin{align}
	k(z,\,s)
	\ &= \
	\int_{-\infty}^{+\infty} \D s'
	\int_{-\infty}^{+\infty} \D z' \
	\kappa(s', z') ,
\end{align}
unsurprisingly, yields the expected solution,
\begin{align}
	\tilde{\tau}(z, \, s)
	\ &= \
	\bs{\chi}^{\T} \bm{K}_{\chi}^{-1} \int_{z}^{s} \D s' \ \kappa(\bs{a}, s') ,
\label{eq:optical_depth_app}
\end{align}
with the corresponding uncertainty estimate given by,
\begin{equation}
\begin{split}
    \tilde{\varepsilon}_{\tau}^{2} (z,\,s)
    \ &= \
	\int_{z}^{s} \D s'
	\int_{z}^{s} \D z' \\
	& \ \ \ \ \ \times
	\left(
	\kappa(s',\,z')
	-
	\kappa(\bs{a},\,s')^{\T}
	\bm{K}_{\chi}^{-1}
    \kappa(\bs{a},\,z')
	\right) .
\end{split}
\label{eq:optdepth_std}
\end{equation}
This can now be used to define the Green's function (\ref{eq:Green}).

It should be emphasised that the uncertainty on the optical depth, and by extension the uncertainty on the opacity, is not yet included in the uncertainty estimate for the radiation field (\ref{eq:characteristics_std}).
The expression (\ref{eq:characteristics_std}) only includes the uncertainties on the emissivity, and not on the opacity or optical depth, because the opacity (only) appears in the linear operator (\ref{eq:RTE_DO}), which in the Bayesian linear regression method is assumed to be deterministic.
The reason for this is that imposing a probability distribution also on the linear operator would render the posterior distribution non-Gaussian, which would severely complicate conditioning and impede analytical solutions.

Nevertheless, in this particular case, an analytic solution can still be obtained for the expectation and the variance of the radiation field, taking into account the distribution of the opacity, although the resulting distribution is not a Gaussian anymore.

From equations (\ref{eq:characteristics_prob}) and (\ref{eq:characteristics_std}), and equations (\ref{eq:optical_depth_app}) and (\ref{eq:optdepth_std}), we know the distributions of the stochastic functions $I$ and $\tau$,
\begin{align}
    p\left( I \, | \, \tau \right)
    \ = \
    \mathcal{N}\left(\tilde{I}, \, \tilde{\varepsilon}^{2}_{I}\right) , \\
    p\left( \tau \right)
    \ = \
    \mathcal{N}\left(\tilde{\tau}, \, \tilde{\varepsilon}^{2}_{\tau}\right) .
    \label{eq:ptau}
\end{align}
The expectation, $\hat{I}$, and variance, $\hat{\varepsilon}_{I}^{2}$, of the radiation field with respect to the joint distribution with $\tau$ can then be obtained using the law of total expectation (see Appendix \ref{subsec:tot_exp}),
\begin{align}
    \hat{I} \ \equiv \ \mathbb{E}\left[ I \right]
    \ &= \
    \mathbb{E}_{\tau}\left[ \mathbb{E}\left[ I \, | \, \tau \right] \right] \ = \ \mathbb{E}_{\tau}\left[ \tilde{I} \right] ,
\end{align}
and similarly, using the law of total variance (see Appendix \ref{subsec:tot_var}),
\begin{equation}
\begin{split}
    \hat{\varepsilon}_{I}^{2} \ \equiv \ \mathbb{V}\left[ I \right]
    \ &= \
    \mathbb{E}_{\tau}\left[ \mathbb{V}\left[ I \, | \, \tau \right] \right]
    \ + \ \mathbb{V}_{\tau}\left[ \mathbb{E}\left[ I \, | \, \tau \right] \right] \\
    \ &= \ \mathbb{E}_{\tau}\left[  \tilde{\varepsilon}_{I}^{2} \right]
    \ + \ \mathbb{E}_{\tau}\left[ \tilde{I}^{2} \right]
    \ - \ \hat{I}^{2} .
\end{split}
\end{equation}
Evaluating these expectations yields (see Appendix \ref{subsec:tot}),
\begin{equation}
	\hat{I} (s)
    \ = \
	I_{0} \, e^{-\hat{\tau}(s_{0},s)}
	\ + \
	\bs{\eta}^{\T} \, \bm{K}_{\eta}^{-1}
    \int_{s_{0}}^{s} \D s' \
    \kappa(\bs{a}, s') \
    e^{-\hat{\tau}(s',s)}
\label{eq:tot_I_exp}
\end{equation}
with the corresponding uncertainty estimate given by,
\begin{equation}
\begin{split}
    \hat{\varepsilon}_{I}^{2} (s)
    \ &\leq \
	\int_{s_{0}}^{s} \D s'
	\int_{s_{0}}^{s} \D z' \
	e^{-\overline{\tau}(s', \, s)}
	e^{-\overline{\tau}(z', \, s)} \\
	& \ \ \ \ \ \times
	\left(
	\kappa(s',\,z')
	-
	\kappa(\bs{a},\,s')^{\T}
	\bm{K}_{\eta}^{-1}
    \kappa(\bs{a},\,z')
	\right) \\
	& \ \ \ \ \ + \ \bar{I}^{2}(s) \ - \ \hat{I}^{2}(s) .
\end{split}
\label{eq:tot_I_var}
\end{equation}
These expressions look very similar to equations (\ref{eq:characteristics_prob}) and (\ref{eq:characteristics_std}). The only difference is that the optical depth is replaced by newly defined effective optical depths,
\begin{align}
\hat{\tau}(z,s) \ &\equiv \ \tilde{\tau}(z,s) \ - \ \frac{1}{2} \, \tilde{\varepsilon}_{\tau}^{2}(z,s) ,
\label{eq:optdepth_eff} \\
\overline{\tau}(z,s) \ &\equiv \ \hat{\tau}(z,s) \ - \ \frac{1}{2} \, \tilde{\varepsilon}_{\tau}^{2}(z,s) ,
\label{eq:optdepth_eff2}
\end{align}
and there are additional terms in (\ref{eq:tot_I_exp}), which account for correlations in the optical depth.
The intensity, $\bar{I}$, is defined analogously to $\hat{I}$, but with $\hat{\tau}$ replaced by $\overline{\tau}$.
Equation (\ref{eq:tot_I_var}) only gives a practical upper bound.
In Appendix \ref{subsec:tot}, we also derive the complete expression for $\hat{\varepsilon}_{I}^{2}$.
The uncertainty on the optical depth thus causes an effective reduction of the optical depth that appears in the radiation field (\ref{eq:tot_I_exp}).

%%%%%%%%%%%%%%%%%%%%
\subsection{Example}
%%%%%%%%%%%%%%%%%%%%
We illustrate the Bayesian method of characteristics with a simple example.
Consider a set of $N_{\text{d}}$ points, $\{s_{d}\}$, at which we know the emissivities, $\{\eta_{d}\}$, and opacities, $\{\chi_{d}\}$. Moreover, consider a set of $N_{\text{b}} = N_{\text{d}}$ basis functions that satisfy the interpolation property for the data points, i.e. $\phi_{i}(s_{d})=\delta_{di}$, such that the design matrix is an identity matrix.
As a result, we have that $\kappa(\bs{a}, \bs{a}) = \bs{\lambda}^{2}$, such that, if we use the same interpolation scheme both for $\eta$ and $\chi$, we have that, $\bm{K}_{\eta} \equiv \bs{\lambda}^{2} + \bs{\sigma}_{\eta}^{2}$, and $\bm{K}_{\chi} \equiv \bs{\lambda}^{2} + \bs{\sigma}_{\chi}^{2}$.
Furthermore, one can show that $\kappa(\bs{a}, s) = \bs{\lambda}^{2} \bs{\phi}(s)$.
Finally, we assume that $\bs{\lambda} = \lambda \mathbb{1}$, and we assume the same uncertainty for every data point, such that $\bs{\sigma}_{\eta} = \sigma_{\eta} \mathbb{1}$ and $\bs{\sigma}_{\chi} = \sigma_{\chi} \mathbb{1}$.
The resulting optical depth and the corresponding uncertainty estimate can then be written as,
\begin{align}
	\tilde{\tau}(z, \, s)
	\ &= \
	\frac{\lambda^{2}}{\lambda^{2} + \sigma_{\chi}^{2}} \,
	\bs{\chi}^{\T}
	\bs{\psi}_{\tau}(z, s) , \\
    \tilde{\varepsilon}_{\tau}^{2} (z,\,s)
    \ &= \
	\frac{\lambda^{2} \sigma_{\chi}^{2}}{\lambda^{2} + \sigma_{\chi}^{2}} \,
    \bs{\psi}_{\tau}(z, s)^{\T} \, \bs{\psi}_{\tau}(z, s) ,
\end{align}
in which $\bs{\chi}$ is the vector of opacities, and we defined the vector,
\begin{align}
	\bs{\psi}_{\tau}(z, s)
	\ \equiv \
	\int_{z}^{s} \D s' \ \bs{\phi}(s') .
\end{align}
Similarly for the radiation field, we find that,
\begin{align}
	\hat{I} (s)
    \ &= \
	I_{0} \, e^{-\hat{\tau}(s_{0},s)} \ + \
	\frac{\lambda^{2}}{\lambda^{2} + \sigma_{\eta}^{2}} \,
	\bs{\eta}^{\T}
	\bs{\hat{\psi}}_{I}(s) , \\
    \hat{\varepsilon}_{I}^{2} (s)
    \ &\leq \
	\frac{\lambda^{2} \sigma_{\eta}^{2}}{\lambda^{2} + \sigma_{\eta}^{2}} \,
    \bs{\bar{\psi}}_{I}(s)^{\T} \, \bs{\bar{\psi}}_{I}(s) \ + \ \bar{I}^{2} (s)  \ - \ \hat{I}^{2} (s),
    \label{eq:epsapp}
\end{align}
in which $\bs{\eta}$ is the vector of emissivities, and we defined the vectors,
\begin{align}
	\bs{\hat{\psi}}_{I}(s)
	\ \equiv \
	\int_{s_{0}}^{s} \D s' \ \bs{\phi}(s') \, e^{-\hat{\tau}(s', \, s)} , \\
	\bs{\bar{\psi}}_{I}(s)
	\ \equiv \
	\int_{s_{0}}^{s} \D s' \ \bs{\phi}(s') \, e^{-\overline{\tau}(s', \, s)} .
\end{align}

\begin{figure}
\centering
\includegraphics[width=0.9\columnwidth]{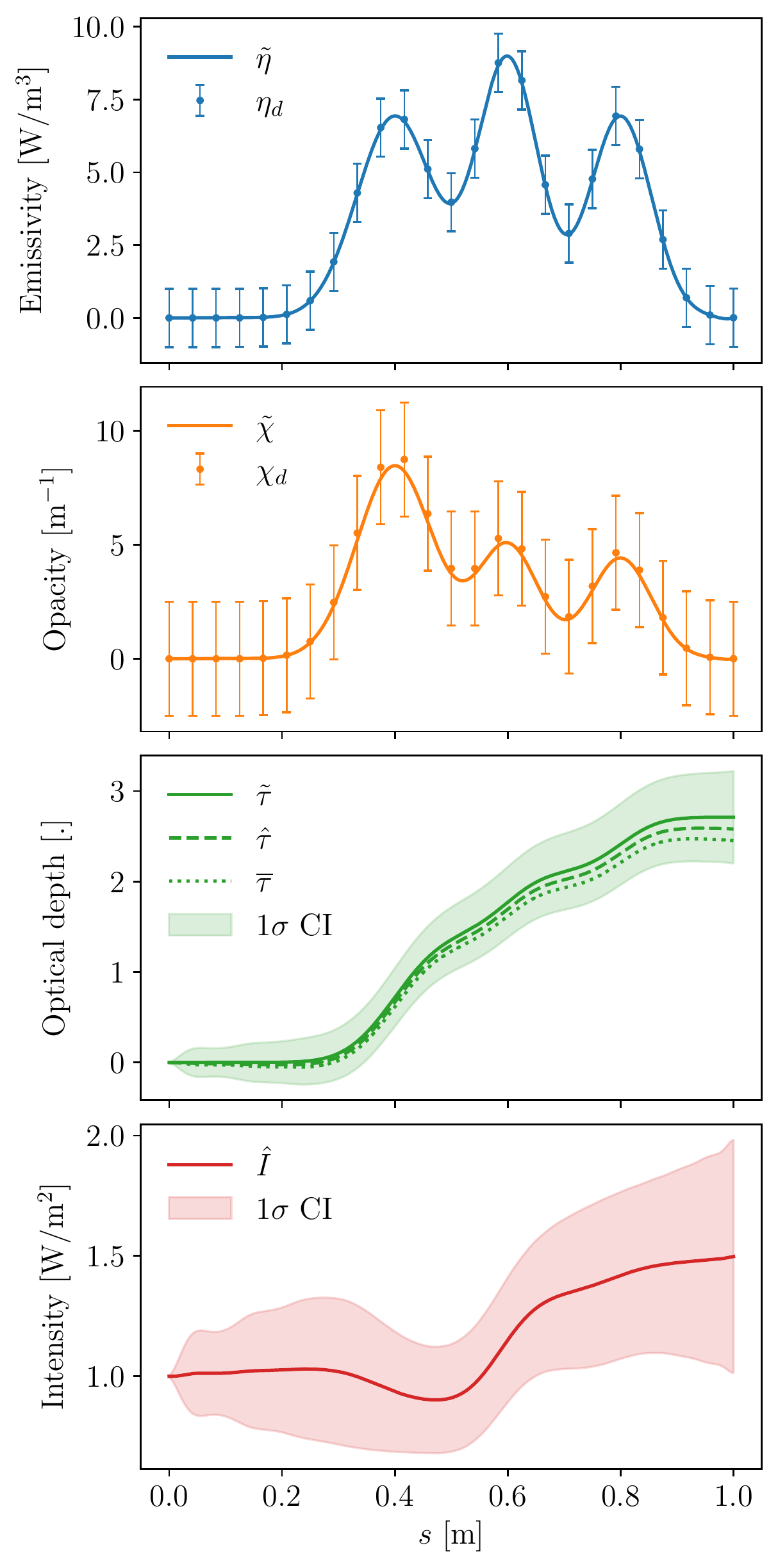}
\caption{Example of the Bayesian method of characteristics for 25 data points, with $I_{0}=1.0$ W/m$^{2}$, $\sigma_{\eta}=1.0$ W/m$^{3}$, $\sigma_{\chi}=2.5$ m$^{-1}$, and $\lambda=10.0$. The error bars on the data and the shaded areas around the curves show the respective (local) 1$\sigma$ confidence intervals (CI), or its upper bound (\ref{eq:epsapp}) for the intensity. The source code for this figure can be found at \href{https://github.com/FredDeCeuster/RadiativeTransferAsRegression}{github.com/FredDeCeuster/RadiativeTransferAsRegression}.}
\label{fig:example}
\end{figure}

Figure \ref{fig:example} shows the solution the the radiative transfer equation along a single ray using the Bayesian method of characteristics.
The variances were chosen comically large, especially to illustrate the effective optical depths.
Even with such large variance, we see that the effect on the optical depth is relatively small.
As basis functions, we choose the 25 fifth-order basis splines that interpolate the 25 uniformly distributed data points, i.e. such that $\phi_{i}(s_{d})=\delta_{di}$.
It should be emphasised that, although the probability distributions for the emissivity, opacity, and optical depth are all Gaussian, the probability distribution for the intensity is not a Gaussian.
In fact, we have not specified the particular distribution, but could nevertheless determine the expectation and variance.

Although we only presented the probabilistic numerical method in the absence of scattering, neglecting any frequency dependence, and only along a single ray, we should note that it can readily by generalised to a three dimensions, including scattering and frequency dependence.
How this can be done for a particular set of basis functions will be demonstrated in a forthcoming paper.

%%%%%%%%%%%%%%%%%%%%%%%%%%%%%%%%%%%
\section{Discussion}
\label{sec:discussion}
%%%%%%%%%%%%%%%%%%%%%%%%%%%%%%%%%%%
The probabilistic numerical method presented in this paper differs significantly from commonly-used (probabilistic) sampling-based Monte Carlo methods for uncertainty quantification \citep[][]{Metropolis1949}.
Where sampling-based Monte Carlo methods are non-intrusive and treat the physical model under consideration as a black box, the probabilistic numerical approach, as described here, requires to recast the entire description of the model as a Bayesian linear regression problem.
This investment, however, pays off in a key advantage for the probabilistic numerical approach:
where sampling-based methods typically require many model evaluations to obtain a distribution, the probabilistic numerical approach requires only a single, albeit computationally slightly more expensive, model evaluation.
This is particularly advantageous for computationally expensive models, such as the ones encountered in radiative transfer, as described in this paper.

The (Bayesian) linear regression model is critically defined by the choice of basis functions or the choice of the corresponding kernel. 
As discussed in Section \ref{subsec:examples}, different choices of basis functions give rise to different kinds of descriptions, or, when, $N_{\text{b}} < N_{\text{d}}$, different kinds of approximations.
In particular, the truncation of a set of Fourier basis functions implies a characteristic minimal resolvable length scale, whether or not to include certain data-centred radial basis functions will alter the solution around those data points, and wavelets allow one to locally refine or coarsen the model.
All of these particular bases have their particular advantages and disadvantages, but none of them is in any sense optimal.
Furthermore, it should be noted that, in our discussion of different bases, we did not take into account the effect of the operator acting on the basis functions, while at the end of Section \ref{subsubsec:PDE} it was emphasised that to solve a linear operator equation such as (\ref{eq:PDE}), the basis functions should ideally be chosen such that $\{\mathscr{L}\phi_{i}\}$ can properly fit $g(x)$, $\{\mathscr{B}\phi_{i}\}$ can properly fit $h(x)$, and $\{\phi_{i}\}$ can properly fit the sought after solution function $f(x)$.
There are many different ways to solve this optimisation problem of finding an appropriate (reduced) set of basis functions, often colloquially referred to as model-order reduction methods \citep[see e.g.][]{Benner2017}. 
One particularly interesting method by \cite{Owhadi2017} describes how to construct a basis that is in a sense optimal, based on probabilistic numerical considerations \citep[see also][]{Owhadi2019}.
In a forthcoming paper, we will choose a particular type of basis, show how it can be tailored to the problem at hand, and demonstrate how this can be used to solve radiative transfer problems in a practical three-dimensional setting.

%%%%%%%%%%%%%%%%%%%%%%%%%%%%%%%%%%%%%%%%%
% \subsubsection{Beyond radiative transfer}
%%%%%%%%%%%%%%%%%%%%%%%%%%%%%%%%%%%%%%%%%
Probabilistic numerical methods are by no means restricted to radiative transfer applications and can readily be applied to various other solvers of operator equations.
In particular, we envision similar techniques to be useful, for instance, in chemical kinetics models that simulate the chemical evolution of a set of species, given a network of chemical reactions \citep[see e.g.][]{McElroy2013}.
These chemical networks are also often reduced to lower the computational cost \citep[see e.g.][]{Grassi2021}.
As a result, the probabilistic numerical setting might also there lead to interesting approximation techniques.
However, since these are initial value problems, the required probabilistic numerical approach will probably be different from what was presented here and more along the lines of, for instance, \cite{Conrad2017}.

%%%%%%%%%%%%%%%%%%%%%%%%%%%%%%%%%%%
\section{Conclusion}
\label{sec:conclusion}
%%%%%%%%%%%%%%%%%%%%%%%%%%%%%%%%%%%
% %The last numbered section should briefly summarise what has been done, and describe
% %the final conclusions which the authors draw from their work.
Inspired by the probabilistic numerical approaches advocated, amongst others, by \cite{Henning2015, Hennig2022} and \cite{Cockayne2019}, we have presented a way to view radiative transfer as a Bayesian linear regression problem.
Specifically, we have modelled the solution of a radiative transfer problem with the expectation of a multivariate Gaussian probability distribution over possible solutions, conditioned on evaluations of the radiative transfer equation and boundary conditions.
This allowed us to model uncertainties, both on the input and output of the model, with the variances of the associated probability distributions, without the need for computationally expensive (Monte Carlo) sampling schemes. 
Moreover, this method naturally allowed us to create reduced-order radiative transfer models, for which the probabilistic interpretation furthermore allowed us to quantify the uncertainty that was introduced by reducing the model.
As an example, we showed how the commonly-used method of characteristics can be derived from a probabilistic point of view.

The aim of this paper was not to present the definitive probabilistic numerical approach for radiative transfer, but rather to motivate future research in this direction by showing the potential benefits of a probabilistic point of view and indicate connections with other research, for instance, by quantifying uncertainties from a statistical point or view, and viewing model reduction as a form of data compression.

%%%%%%%%%%%%%%%%%%%%%%%%%%%%%%%%%%%%%%%%%%%%%%%%%%
\section*{Acknowledgements}
%%%%%%%%%%%%%%%%%%%%%%%%%%%%%%%%%%%%%%%%%%%%%%%%%%
We would like to thank Maarten Baes and Peter Camps for planting the idea that the future of radiative transfer probably lies somewhere in between the typical probabilistic and deterministic approaches.
We also thank Amery Gration and Johan Suykens for insightful discussions on Bayesian linear regression and Gaussian processes.
Finally, we would also like to thank the scientific editor and the two anonymous reviewers for their considerate and helpful feedback.
FDC is supported by the EPSRC iCASE studentship programme (ref. 1878976) and Intel Corporation. FDC and LD acknowledge support from the ERC consolidator grant 646758 AEROSOL.
TC is a PhD fellow of the Research Foundation - Flanders (FWO).

%%%%%%%%%%%%%%%%%%%%%%%%%%%%%%%%%%%%%%%%%%%%%%%%%%
\section*{Data availability}
%%%%%%%%%%%%%%%%%%%%%%%%%%%%%%%%%%%%%%%%%%%%%%%%%%
No new data were generated or analysed in support of this research.

%%%%%%%%%%%%%%%%%%%%%%%%%%%%%%%%%%%%%%%%%%%%%%%%%%


\bibliographystyle{mnras}
\bibliography{references}

%\begin{thebibliography}{99}
%\end{thebibliography}

%%%%%%%%%%%%%%%%%%%%%%%%%%%%%%%%%%%%%%%%%%%%%%%%%%

%%%%%%%%%%%%%%%%% APPENDICES %%%%%%%%%%%%%%%%%%%%%

\appendix

\section{Mathematical background}

% This appendix provides some mathematical background supporting the claims in the main text that were presented without proof.

%%%%%%%%%%%%%%%%%%%%%%%%%%%%%%%%%%%%%%%%%%%%%%%%%%%%%%%%%%%%
\subsection{Equivalence between primal and dual formulation}
\label{appendix:duality}
%%%%%%%%%%%%%%%%%%%%%%%%%%%%%%%%%%%%%%%%%%%%%%%%%%%%%%%%%%%%
To show the equivalence between the primal and dual formulation we have to prove equation (\ref{eq:duality}), or equivalently,
\begin{equation}
    \bs{\sigma}^{-2} \bs{\Phi}
	\left(
	    \bs{\Phi}^{\T} \bs{\sigma}^{-2} \bs{\Phi}
	   	+
	    \bs{\lambda}^{-2}
	\right)^{-1}
	\ - \
	\left(
	    \bs{\Phi} \bs{\lambda}^{2} \bs{\Phi}^{\T}
	    +
	    \bs{\sigma}^{2}
	\right)^{-1}
	\bs{\Phi} \bs{\lambda}^{2}
	\ = \
	0 .
\label{eq:A1}
\end{equation}
Using the Woodburry matrix identity, the second term expands as,
\begin{equation}
    \bs{\sigma}^{-2} \bs{\Phi} \bs{\lambda}^{2}
    -
    \bs{\sigma}^{-2} \bs{\Phi}
    \left(
        \bs{\lambda}^{-2}
        +
        \bs{\Phi}^{\T} \bs{\sigma}^{-2} \bs{\Phi}
    \right)^{-1}
    \bs{\Phi}^{\T} \bs{\sigma}^{-2}
    \bs{\Phi} \bs{\lambda}^{2} .
\label{eq:A2}
\end{equation}
Using (\ref{eq:A2}) in (\ref{eq:A1}), ignoring the overall factor, $\bs{\sigma}^{-2} \bs{\Phi}$, and isolating the terms with the inverse, it remains to show that,
\begin{equation}
	\left(
	    \bs{\Phi}^{\T} \bs{\sigma}^{-2} \bs{\Phi}
	   	+
	    \bs{\lambda}^{-2}
	\right)^{-1}
	\left(
	\mathbb{1}
	+
    \bs{\Phi}^{\T} \bs{\sigma}^{-2}
    \bs{\Phi} \bs{\lambda}^{2}
	\right)
	-
    \bs{\lambda}^{2}
    \ = \
    0 .
\end{equation}
Rewriting the second factor by extracting $\bs{\lambda}^{2}$ then yields,
\begin{equation}
	\left(
	    \bs{\Phi}^{\T} \bs{\sigma}^{-2} \bs{\Phi}
	   	+
	    \bs{\lambda}^{-2}
	\right)^{-1}
	\left(
    \bs{\lambda}^{-2}
	+
    \bs{\Phi}^{\T} \bs{\sigma}^{-2}
    \bs{\Phi}
	\right)
    \bs{\lambda}^{2}
	-
    \bs{\lambda}^{2}
    \ = \
    0
\end{equation}
making it clear that the equality indeed holds and that the primal and dual solutions are thus equivalent.

%%%%%%%%%%%%%%%%%%%%%%%%
\subsection{Definitions}
\label{subsec:defs}
%%%%%%%%%%%%%%%%%%%%%%%%
In this section, we summarise and explain some of the concepts from statistics that are used throughout the main text.

%%%%%%%%%%%%%%%%%%%%%%%%%%%
\subsubsection{Expectation}
%%%%%%%%%%%%%%%%%%%%%%%%%%%
The expectation, $\mathbb{E}[X]$, of a random variable, $X$, is defined as the integral (or sum) over all possible values, $x$, that variable can take, weighted by its probability density, $p(x)$,
\begin{equation}
    \mathbb{E}[X] \ \equiv \ \int \D x \ p(x) \, x. 
\end{equation}
Although here, in our notation, we carefully distinguished between the random variable, $X$, and a specific realisation of that variable, $x$, throughout this paper, we sometimes make the common slight abuse of notation by denoting both a random variable and its realisations with the same symbol.

%%%%%%%%%%%%%%%%%%%%%%%%
\subsubsection{Variance}
%%%%%%%%%%%%%%%%%%%%%%%%
The variance, $\mathbb{V}[X]$, of a random variable, $X$, is defined as the expectation of the square difference between that variable and its expectation,
\begin{equation}
    \mathbb{V}[X] \ \equiv \ \mathbb{E}\big[ \left(X - \mathbb{E}[X] \right)^{2} \big] .
\label{eq:def_var}
\end{equation}
The variance can be interpreted as the expected square deviation from its expectation and thus quantifies the spread of the distribution of the random variable.
Sometimes, the variance can be computed more conveniently as,
\begin{equation}
    \mathbb{V}[X] \ = \ \mathbb{E}\big[ X^{2} \big] \ - \ \mathbb{E}\left[ X \right]^{2} ,
\label{eq:var2}
\end{equation}
which follows from the definition (\ref{eq:def_var}) by direct computation.

%%%%%%%%%%%%%%%%%%%%%%%%%%
\subsubsection{Covariance}
%%%%%%%%%%%%%%%%%%%%%%%%%%
The covariance, $\text{Cov}[X_{i}, X_{j}]$, between two random variables, $X_{i}$ and $X_{j}$, is defined as, the expectation of the product of the differences of each variable with its expectation,
\begin{equation}
    \text{Cov}[X_{i}, X_{j}] \ \equiv \ \mathbb{E}\big[ \left(X_{i} - \mathbb{E}[X_{i}] \right)\left(X_{j} - \mathbb{E}[X_{j}] \right) \big] .
\label{eq:def_cov}
\end{equation}
The covariance of a random variable and itself equals its variance,
\begin{equation}
    \text{Cov}[X_{i}, X_{i}] \ = \ \mathbb{V}[X_{i}] ,
\end{equation}
which follows directly from the definitions (\ref{eq:def_var}) and (\ref{eq:def_cov}).

%%%%%%%%%%%%%%%%%%%%%%%%%%%%%%%%%%%%
\subsubsection{Marginal probability}
%%%%%%%%%%%%%%%%%%%%%%%%%%%%%%%%%%%%
The marginal probability distribution, $p(X_{i})$, of a random variable, $X_{i}$, given the joint probability distribution, $p(X_{i}, X_{j})$, with another random variable, $X_{j}$, is given by,
\begin{equation}
    p(X_{i}) \ \equiv \ \int \D X_{j} \ p(X_{i}, X_{j}) ,
\end{equation}
which amounts to integrating out the other random variable.
Note the abuse of notation in using the random variable, $X_{j}$, to denote its observed value, $x_{j}$.

%%%%%%%%%%%%%%%%%%%%%%%%%%%%%%%%%%%%%%%
\subsubsection{Conditional probability}
%%%%%%%%%%%%%%%%%%%%%%%%%%%%%%%%%%%%%%%
The conditional probability, $p(X_{i} | x_{j})$, of a random variable, $X_{i}$, given the observation of the value, $x_{j}$, of another random variable, $X_{j}$, is given by,
\begin{equation}
    p(X_{i} | X_{j}) \ \equiv \ \frac{p(X_{i}, X_{j})}{\int \D X_{i} \ p(X_{i}, X_{j})} \ = \ \frac{p(X_{i}, X_{j})}{p(X_{j})} ,
\end{equation}
which amounts to a re-scaling of the joint distribution, $p(X_{i}, X_{j})$, with the marginal distribution $p(X_{j})$.
Note the abuse of notation in using the random variable, $X_{j}$, to denote its observed value, $x_{j}$.

%%%%%%%%%%%%%%%%%%%%%%%%%%%%%%%%%%%%%%%%%%%%%%%%%%%%%%%%%%%%%%%%%%%%%%%%
\subsubsection{Multivariate Gaussian or normal probability distribution}
%%%%%%%%%%%%%%%%%%%%%%%%%%%%%%%%%%%%%%%%%%%%%%%%%%%%%%%%%%%%%%%%%%%%%%%%
A random vector variable, $\bs{X} \sim \mathcal{N}(\bs{\mu}, \bs{\Sigma})$, follows a multivariate Gaussian or normal probability distribution with mean vector, $\bs{\mu}$, and covariance matrix, $\bs{\Sigma}$, if its probability distribution is given by,
\begin{equation}
    p(X) \ \equiv \ \frac{1}{\sqrt{\det \left( 2 \pi \bs{\Sigma} \right)}} \ \exp \left(-\frac{1}{2} \left(\bs{X}-\bs{\mu} \right)^{\text{T}} \bs{\Sigma}^{-1} \left(\bs{X}-\bs{\mu} \right) \right) .
\end{equation}
By direct computation, one can verify that the expectation, variance, and covariance of the components, $X_{i}$, of the multivariate Gaussian distributed vector variable $\bs{X} \sim \mathcal{N}(\bs{\mu}, \bs{\Sigma})$, are respectively given by,
\begin{align}
    \mathbb{E}[X_{i}]        \ &= \ \mu_{i}      , \\
    \mathbb{V}[X_{i}]        \ &= \ \Sigma_{i i} , \\ 
    \text{Cov}[X_{i}, X_{j}] \ &= \ \Sigma_{i j} .
\end{align}
The relations between a marginal and conditional (multivariate) Gaussian distributions are given in Appendix \ref{app:margcondGauss} and the relations for conditioning a (multivariate) Gaussian distribution are given in Appendix \ref{app:condGauss}.

%%%%%%%%%%%%%%%%%%%%%%%%%%%%%%%%%%%%%%%%%%%%%%
\subsection{Marginal \& conditional Gaussians}
\label{app:margcondGauss}
%%%%%%%%%%%%%%%%%%%%%%%%%%%%%%%%%%%%%%%%%%%%%%
Given a (marginal) Gaussian distribution, $p(\bs{x})$, and a corresponding conditional Gaussian distribution,
$p(\bs{y}\,|\,\bs{x})$, which are defined as,
\begin{align}
    p(\bs{x})
	\ &= \
	\mathcal{N}
	\big(
		\bs{\mu_{x}}, \ \ \bs{\Sigma}_{\bs{x}}
	\big), \\
    p(\bs{y}\,|\,\bs{x})
	\ &= \
	\mathcal{N}
	\big(
		A \bs{x} + \bs{b}, \ \ \bs{\Sigma}_{\bs{y}|\bs{x}}
	\big),
\end{align}
the other corresponding marginal distribution, $p(\bs{y})$, and the reverse conditional distribution,
$p(\bs{x}\,|\,\bs{y})$, are given by,
\begin{align}
    p(\bs{y})
	\ &= \
	\mathcal{N}
	\left(
		A \bs{\mu_{x}} + \bs{b}, \ \
		\bs{\Sigma}_{\bs{y}|\bs{x}} \, + \, A \bs{\Sigma}_{\bs{x}} A^{\T}
	\right), \\
    p(\bs{x}\,|\,\bs{y})
	\ &= \
	\mathcal{N}
	\bigg(
		\bs{\Sigma}
		\left(
			A^{\T} \bs{\Sigma}^{-1}_{\bs{y}|\bs{x}} (\bs{y}-\bs{b})
			\ + \
			\bs{\Sigma}^{-1}_{\bs{x}} \bs{\mu_{x}}
		\right), \ \
		\bs{\Sigma}
	\bigg),
\end{align}
in which we defined the covariance matrix,
\begin{equation}
	\bs{\Sigma}
	\ \equiv \
	\left(
		\bs{\Sigma}^{-1}_{\bs{x}} \, + \, A^{\T} \bs{\Sigma}^{-1}_{\bs{y}|\bs{x}} A
	\right)^{-1} .
\end{equation}
These relations can be derived by ``completing the square'' in the distribution function and collecting the relevant terms, as described in detail, for instance, in \cite{Bishop2006}.

%%%%%%%%%%%%%%%%%%%%%%%%%%%%%%%%%%%%
\subsection{Conditioning a Gaussian}
\label{app:condGauss}
%%%%%%%%%%%%%%%%%%%%%%%%%%%%%%%%%%%%
Consider a stochastic vector variable, $\bs{y}$, defined by two separate stochastic vector variables, $\bs{a}$ and $\bs{b}$, and assume that all components follow a (multivariate) Gaussian distribution, i.e.,
\begin{equation}
	\bs{y}
	\ = \
	\left[
		\begin{matrix}
			\bs{a} \\
			\bs{b}
		\end{matrix} 
	\right]\
	\ \sim \
	\mathcal{N}
	\left(
        \left[
		    \begin{matrix}
	    	    \bs{\mu}_{\bs{a}} \\
	    	    \bs{\mu}_{\bs{b}}
     	    \end{matrix}
	    \right],
        \left[
    	\begin{matrix}
		    \bs{\Sigma}_{\bs{a} \bs{a}} & \bs{\Sigma}_{\bs{a} \bs{b}} \\
            \bs{\Sigma}_{\bs{b} \bs{b}} & \bs{\Sigma}_{\bs{b} \bs{b}}
	    \end{matrix}
	    \right]
	\right) ,
\end{equation}
in which, $\bs{\mu}_{\bs{a}}$ and $\bs{\mu}_{\bs{b}}$ are the mean vectors and the matrices $\bs{\Sigma}_{\bs{a} \bs{a}}$,
$\bs{\Sigma}_{\bs{a} \bs{b}} = \bs{\Sigma}_{\bs{b} \bs{a}}^{\T}$, and $\bs{\Sigma}_{\bs{b} \bs{b}}$, together form the covariance matrix.
Now, we can ask what the resulting distribution of $\bs{a}$ would be, given prior knowledge about the value for $\bs{b}$.
Fixing the value of $\bs{b}$ again yields a multivariate Gaussian distribution,
\begin{equation}
	p\left(\bs{a}	 \ | \ \bs{b} \right)
	\ = \
	\mathcal{N}
    \big(
	    \bs{\mu}_{\bs{a} | \bs{b}},
        \bs{\Sigma}_{\bs{a} | \bs{b}}
    \big),
	\end{equation}
in which the conditioned mean and variance are given by,
\begin{align}
    \bs{\mu}_{\bs{a} | \bs{b}}
	\ &= \
	\bs{\mu}_{\bs{a}} \ + \ \bs{\Sigma}_{\bs{a} \bs{b}} \, \bs{\Sigma}_{\bs{b} \bs{b}}^{-1}
	\left( \bs{b} - \bs{\mu}_{\bs{b}} \right), \\
    \bs{\Sigma}_{\bs{a} | \bs{b}}
	\ &= \
	\bs{\Sigma}_{\bs{a} \bs{a}} \ - \ \bs{\Sigma}_{\bs{a} \bs{b}} \, \bs{\Sigma}_{\bs{b} \bs{b}}^{-1} \, \bs{\Sigma}_{\bs{b} \bs{a}}.
\end{align}
These relations can be derived by ``completing the square'' in the distribution function and collecting the relevant terms, as described in detail, for instance, in \cite{Bishop2006}.

Note that without correlation between $\bs{a}$ and $\bs{b}$, i.e. when $\bs{\Sigma}_{\bs{a} \bs{b}} =
\bs{\Sigma}_{\bs{b} \bs{a}}^{\T} = 0$, the prior knowledge about $\bs{b}$ will not affect the
distribution of $\bs{a}$, which is in line with expectations.

%%%%%%%%%%%%%%%%%%%%%%%%%%%%%%%%%%%%%%%%%%%%%%%%
\subsection{RKHS bound on the uncertainty}
\label{app:RKHSbound}
%%%%%%%%%%%%%%%%%%%%%%%%%%%%%%%%%%%%%%%%%%%%%%%%
Let $\mathcal{H}$ denote the reproducing kernel Hilbert space (RKHS) of the kernel defined in (\ref{eq:kernel}), with an associated inner product, $\langle \rangle_{\mathcal{H}}$, and norm $\|\cdot\|_{\mathcal{H}} \equiv \sqrt{\langle \cdot, \cdot\rangle_{\mathcal{H}}}$.
The defining properties of an RKHS with reproducing kernel, $k$, are that, $k(x, \cdot) \in \mathcal{H}$, and that,
\begin{equation}
\forall f \in \mathcal{H} : \ \left\langle f,  k(x, \cdot) \right\rangle_{\mathcal{H}} \ = \ f(x),
\label{eq:rep1}
\end{equation}
i.e. an inner product with the kernel around $x$ corresponds to function evaluation in $x$
\citep[see e.g.][for a comprehensive introduction]{Berlinet2004}.
The latter is known as the reproducing property and it is the key to derive the bound given in equation (\ref{eq:RKHSbound}). 
If we now consider the projection $Pf\in\mathcal{H}$ of the function, $f$, in the RKHS, $\mathcal{H}$, the reproducing property (\ref{eq:rep1}) implies that,
\begin{equation}
Pf(x) \ = \ \left\langle Pf,  k(x, \cdot) \right\rangle_{\mathcal{H}} .
\label{eq:rep}
\end{equation}
By definition of the data, we also have that $Pf(\x)=\y$, such that,
\begin{equation}
\y \ = \ \left\langle Pf,  k(\x, \cdot) \right\rangle_{\mathcal{H}} .
\label{eq:datarep}
\end{equation}
Substituting this in (\ref{eq:mean_dual}), and defining $\bm{K} \equiv k(\x, \x) + \bs{\sigma}^{2}$, yields,
\begin{equation}
\tilde{f}(x)
    \ = \
    k(x, \bs{x}) \, \bm{K}^{-1} \left\langle Pf,  k(\x, \cdot) \right\rangle_{\mathcal{H}} ,
\label{eq:above}
\end{equation}
such that, in combination with equation (\ref{eq:rep}), we find that,
\begin{equation}
\begin{split}
    &\left| Pf(x) \ - \ \tilde{f}(x) \right| \\
    &\ = \
    \left| \left\langle Pf, \, k(\cdot, x) - k(x, \bs{x}) \, \bm{K}^{-1}  k(\cdot, \x) \right\rangle_{\mathcal{H}} \right| \\
    &\ \leq \
    \left\| Pf \right\|_{\mathcal{H}} \, \left\| k(x, \cdot) - k(x, \bs{x}) \, \bm{K}^{-1}  k(\x, \cdot) \right\|_{\mathcal{H}},
\end{split}
\label{eq:bound}
\end{equation}
where in the last step, we used the Cauchy-Schwarz inequality.
Considering the square of the last factor, we find,
\begin{equation}
\begin{split}
    &\left\| k(x, \cdot) - k(x, \bs{x}) \, \bm{K}^{-1}  k(\x, \cdot) \right\|_{\mathcal{H}}^{2} \\
    & \ = \
    k(x,x) \ - \ 2 k(x, \bs{x}) \, \bm{K}^{-1}  k(\x, x) \\
    & \ \ \ \ \ + \ k(x, \bs{x}) \, \bm{K}^{-1}  k(\x, \x) \, \bm{K}^{-1}  k(\x, x) \\
    & \ = \
    k(x,x) \ - \ k(x, \bs{x}) \, \bm{K}^{-1}  k(\x, x) \\
    & \ \ \ \ \ - k(x, \bs{x}) \, \bm{K}^{-1}  \big( \bm{K} - k(\x, \x) \big) \, \bm{K}^{-1}  k(\x, x) \\
    & \ = \
    \tilde{\varepsilon}^{2}(x)
    \ - \ k(x, \bs{x}) \, \bm{K}^{-1} \bs{\sigma}^{2} \, \bm{K}^{-1}  k(\x, x) ,
\end{split}
\end{equation}
where in the last equality we used equation (\ref{eq:std_dual}). Since the second term can be viewed as (minus) the square of the Euclidean norm of the vector, $\bs{\sigma} \, \bm{K}^{-1}  k(\x, x)$, it will always be negative, such that,
\begin{equation}
    \left\| k(x, \cdot) - k(x, \bs{x}) \, \bm{K}^{-1}  k(\x, \cdot) \right\|_{\mathcal{H}}
    \ \leq \
    \tilde{\varepsilon}(x).
\end{equation}
Note that in the limit of perfect data, i.e. $\bs{\sigma}\rightarrow\bs{0}$, the above inequality becomes an equality.
Substituting this in equation (\ref{eq:bound}), we obtain the desired bound on the local error,
\begin{equation}
    \left|Pf(x) - \tilde{f}(x)\right| \ \leq \ \left\|Pf\right\|_{\mathcal{H}} \ \tilde{\varepsilon}(x) .
\end{equation}
It should be emphasised that this only bounds the absolute difference between the approximation and the projection of the true solution in the RKHS, not the absolute difference between the approximation and the true solution itself.
Therefore, the strength of this bound crucially depends on the RKHS, and thus on the particular kernel, or equivalently, on the particular set of basis functions that is used.

%%%%%%%%%%%%%%%%%%%%%%%%%%%%%%%%%%%%%%%%%%%%%%%%
\subsection{Equivalent kernel for the method of characteristics}
\label{appendix:Green}
%%%%%%%%%%%%%%%%%%%%%%%%%%%%%%%%%%%%%%%%%%%%%%%%
Given a linear PDE, and given the corresponding Green's function, $G$, for the differential operator, one can construct a kernel,
\begin{align}
	k(z, s)
	\ &= \
	\int_{-\infty}^{+\infty} \D s'
	\int_{-\infty}^{+\infty} \D z' \
	\kappa(s', z') \
	G(z', z) \
	G(s', s) ,
\label{eq:desired_kernel}
\end{align}
based on another kernel, $\kappa$. For later convenience, we define a new function, $g$, that, using the Green's functions, can be expressed as,
\begin{align}
	g(s, z)
	\ &\equiv \
	\mathscr{L}_{2} k(z, s)
	\ = \
	\int_{-\infty}^{+\infty} \D z' \
	\kappa(s, z') \
	G(z', z), \\
	g(z, s)
	\ &\equiv \
	\mathscr{L}_{1} k(z, s)
	\ = \
	\int_{-\infty}^{+\infty} \D s' \
	\kappa(s', z) \
	G(s', s),
\end{align}
in which the subscript on the differential operator, $\mathscr{L}$, indicates whether it acts on the first or second argument.
Note that both definitions are consistent, since $k$ is symmetric in its arguments. Using the Green's functions again, one can derive,
\begin{align}
	\mathscr{L}_{1} \mathscr{L}_{2} k(z, s)
	\ &= \
	\mathscr{L}_{1} g(s, z)
	\ = \
	\kappa(s, z) , \\
	\mathscr{L}_{2} \mathscr{L}_{1} k(z, s)
	\ &= \
	\mathscr{L}_{2} g(z, s)
	\ = \
	\kappa(s, z) .
\end{align}
When solving the PDE as a Bayesian linear regression problem, the corresponding covariance matrix of the joint distribution, reads,
\begin{align}
    \left(
		\begin{matrix*}[r]
			\mathscr{L}_{1} \mathscr{L}_{2} k(\bs{a}, \bs{a}) &
			\mathscr{L}_{1} \mathscr{B}_{2} k(\bs{a}, \bs{b}) &
			\mathscr{L}_{1}                 k(\bs{a},  s ) \\
			\mathscr{B}_{1} \mathscr{L}_{2} k(\bs{b}, \bs{a}) &
			\mathscr{B}_{1} \mathscr{B}_{2} k(\bs{b}, \bs{b}) &
			\mathscr{B}_{1}                 k(\bs{b},  s ) \\
			\mathscr{L}_{2}                 k(    s , \bs{a}) &
			\mathscr{B}_{2}                 k(    s , \bs{b}) &
			                                k(    s ,  s ) \\
    	\end{matrix*}
	\right) 
\end{align}
and can be simplified using the definitions above to yield,
\begin{align}
    \left(
		\begin{matrix*}[r]
			                           \kappa(\bs{a}, \bs{a}) &
			                \mathscr{B}_{1} g(\bs{a}, \bs{b}) &
			                                g(\bs{a},     s ) \\
                  			\mathscr{B}_{1} g(\bs{a}, \bs{b}) &
			\mathscr{B}_{1} \mathscr{B}_{2} k(\bs{b}, \bs{b}) &
			\mathscr{B}_{1}                 k(\bs{b},     s ) \\
											g(\bs{a},     s ) &
			\mathscr{B}_{2}                 k(    s , \bs{b}) &
			                                k(    s ,     s ) \\
    	\end{matrix*}
	\right) .
\end{align}
The requirement that this matrix is positive semi-definite for all Green's functions, $G$, can be reduced to the condition that,
\begin{align}
	\kappa(\bs{a},\,s)^{\T} \,
	\kappa(\bs{a},\,\bs{a})^{-1} \,
	\kappa(\bs{a},\,z)
	\ &\leq \
	\kappa(s,\,z),
\end{align}
holds for all $s,z \in D$, which is equivalent to the condition that $\kappa$ is a positive semi-definite kernel, as expected.

In the method of characteristics (Section \ref{subsec:characteristics}), we considered the special case where the second kernel, $\kappa$, has the additional property that it cannot correlate the regions $s>s_{0}$ and $s<s_{0}$, i.e.
\begin{equation}
\begin{split}
	\kappa(z,s)
	\ &\equiv \
	\Theta(s_{0}-z) \Theta(s_{0}-s) 
    \kappa(z,s)  \\
	& \quad \ + \
    \Theta(z-s_{0}) \Theta(s-s_{0})
    \kappa(z,s) .
\end{split}
\label{eq:splitcond}
\end{equation}
Using the Green's function from the radiative transfer equation, this implies, for $z \geq b$ and $s \geq b$, that,
\begin{equation}
\begin{split}
    k(z, s)
    \ &= \
	\int_{b}^{z} \D z'
	\int_{b}^{s} \D s' \
	\kappa(z',\,s') \,
    e^{-\tau(z',z)} \, e^{-\tau(s',s)} \\
	\ & \quad \ + \
    k(b,b) \, e^{-\tau(b,z)}  \, e^{-\tau(b,s)} .
\end{split}
\end{equation}
Similarly, this implies, for $z \geq b$ and $s \geq b$, that,
\begin{align}
	g(s, z)
	\ = \
	\int_{b}^{z} \D z' \
	\kappa(s, z') \
	e^{-\tau(z',z)} ,
\end{align}
and in particular that for $s \geq b$, we have that $g(s,b)=0$. As a result, the inverted matrix in equations (\ref{eq:mean_dual2}) and (\ref{eq:std_dual2}) reduces to,
\begin{equation}
\begin{split}
    &
    \left(
		\begin{matrix*}[l]
			\mathscr{L}_{1} \mathscr{L}_{2} k(\bs{a}, \bs{a}) + \bs{\sigma}_{\text{L}}^{2} &
			\mathscr{L}_{1} \mathscr{B}_{2} k(\bs{a},     b ) \\
			\mathscr{B}_{1} \mathscr{L}_{2} k(    b , \bs{a}) &
			\mathscr{B}_{1} \mathscr{B}_{2} k(    b ,     b ) + \sigma_{\text{B}}^{2}      \\
    	\end{matrix*}
	\right) \\
    &= \
    \left(
    	\begin{matrix*}
			\kappa(\bs{a}, \bs{a}) + \bs{\sigma}_{\text{L}}^{2} & g(\bs{a}, b) \\
			g(\bs{a}, b)^{\T}      & k(    b , b) + \sigma_{\text{B}}^{2}      \\
    	\end{matrix*}
	\right) \\
    &= \
    \left(
    	\begin{matrix*}
			\kappa(\bs{a}, \bs{a}) + \bs{\sigma}_{\text{L}}^{2} & \bs{0} \\
			\bs{0}^{\T}      & k(    b , b) + \sigma_{\text{B}}^{2}      \\
    	\end{matrix*}
	\right)
\end{split}
\end{equation}
Define the matrix $\bm{K} \equiv \kappa(\bs{a}, \bs{a}) + \bs{\sigma}_{\text{L}}^{2}$, and let us assume that there is no uncertainty on the boundary condition, i.e. $\sigma_{\text{B}}=0$. The function approximation in the dual formulation then reads,
\begin{equation}
\begin{split}
	\tilde{f}_{\text{dual}} (s)
    \, &= \,
    \left(
    	\begin{matrix}
    		\bs{\eta} \\
			I_{0}
    	\end{matrix}     
   	\right)^{\T}
    \left(
    	\begin{matrix*}
			\bm{K}      & \bs{0} \\
			\bs{0}^{\T} & k(    b , b) \\
    	\end{matrix*}
	\right)^{-1} 
    \left(
    	\begin{matrix}
			g(\bs{a}, s) \\
			k(    b , s)
    	\end{matrix}
    \right) \\
    \, &= \,
    I_{0}  \, e^{-\tau(b,s)}
    + \, 
    \bs{\eta}^{\T} \bm{K}^{-1}
    \int_{b}^{s} \D s' \
	\kappa(\bs{a}, s')
	e^{-\tau(s',s)} .
\end{split}
\end{equation}
We clearly recognise the result from the method of characteristics.
Similarly, the corresponding variance reads,
\begin{equation}
\begin{split}
	\tilde{\sigma}^{2}_{\text{dual}} (s)
	\, &= \,
	k(s,s)
	-
    \left(
    	\begin{matrix}
			g(\bs{a}, s) \\
			k(    b , s)
    	\end{matrix}
    \right)^{\T}
        \left(
    	\begin{matrix*}
			\bm{K}      & \bs{0} \\
			\bs{0}^{\T} & k(    b , b) \\
    	\end{matrix*}
	\right)^{-1} 
    \left(
    	\begin{matrix}
			g(\bs{a}, s) \\
			k(    b , s)
    	\end{matrix}
    \right) \\
    \, &= \,
	\int_{b}^{s} \D s'
	\int_{b}^{s} \D z' \
	e^{-\tau(s', \, s)}
	e^{-\tau(z', \, s)} \\
	& \ \ \ \ \ \times
	\left(
	\kappa(s',\,z')
	-
	\kappa(\bs{a},\,s')^{\T}
	\bm{K}^{-1}
    \kappa(\bs{a},\,z')
	\right) .
\end{split}
\end{equation}
In the parentheses, we recognise the conditioned variance (\ref{eq:std_dual}) that stems from the interpolation of the emissivity (\ref{eq:int_emi}).

%%%%%%%%%%%%%%%%%%%%%%%%%%%%%%%%%%%%%%%%%
\subsection{The law of total expectation}
\label{subsec:tot_exp}
%%%%%%%%%%%%%%%%%%%%%%%%%%%%%%%%%%%%%%%%%
Given two random variables, $X$ and $Y$, in the same probability space, the law of total expectation states that,
\begin{equation}
    \mathbb{E}[X] \ = \ \mathbb{E}_{Y}\big[\mathbb{E}[X | Y]\big] ,
\label{eq:tot_exp}
\end{equation}
i.e. the expectation of $X$ is the same as the expectation over $Y$ of the conditional expectation of $X$ given $Y$.
A sketch for a proof can be derived from the following equalities,
\begin{equation}
\begin{split}
    \mathbb{E}_{Y}\big[\mathbb{E}[X | Y]\big]
    \ &= \ \int \D Y \ p(Y) \int \D X \ p(X|Y) \, X \\
    \ &= \ \int \D Y \int \D X \ p(X,Y) \, X \\
    \ &= \ \int \D X \ p(X) \, X \\
    \ &= \ \mathbb{E}[X] ,
\end{split}
\end{equation}
where in the first equality we used the definition of the expectation, in the second we used the conditional probability, and in the third we used the marginal probability.

%%%%%%%%%%%%%%%%%%%%%%%%%%%%%%%%%%%%%%%%%%%%
\subsection{The law of total total variance}
\label{subsec:tot_var}
%%%%%%%%%%%%%%%%%%%%%%%%%%%%%%%%%%%%%%%%%%%%
Given two random variables, $X$ and $Y$, in the same probability space, the law of total variance states that,
\begin{equation}
    \mathbb{V}[X] \ = \ \mathbb{E}_{Y}\big[\mathbb{V}[X | Y]\big] \ + \ \mathbb{V}_{Y}\big[\mathbb{E}[X | Y]\big],
\end{equation}
i.e. the variance of $X$ is the sum of the expected conditional variance and the variance of the conditional expectation. This follows directly form the law of total expectation (\ref{eq:tot_exp}).
Using the law of total expectation in equation (\ref{eq:var2}) yields,
\begin{equation}
\begin{split}
    \mathbb{V}[X]
    \ &= \ \mathbb{E}_{Y}\Big[\mathbb{E}\big[X^{2} | Y\big]\Big] \ - \ \mathbb{E}_{Y}\big[\mathbb{E}[X | Y]\big] ^{2} \\
    \ &= \ \mathbb{E}_{Y}\Big[\mathbb{V}\big[X | Y\big] \ + \ \mathbb{E}\big[X | Y\big]^{2}\Big] \ - \ \mathbb{E}_{Y}\big[\mathbb{E}[X | Y]\big] ^{2} \\
    \ &= \ \mathbb{E}_{Y}\Big[\mathbb{V}\big[X | Y\big]\Big] \ + \ \mathbb{E}_{Y}\Big[\mathbb{E}\big[X | Y\big]^{2}\Big] \ - \ \mathbb{E}_{Y}\big[\mathbb{E}[X | Y]\big] ^{2} \\
    \ &= \ \mathbb{E}_{Y}\big[\mathbb{V}[X | Y]\big] \ + \ \mathbb{V}_{Y}\big[\mathbb{E}[X | Y]\big],
\end{split}
\end{equation}
where in the second and fourth equality we used equation (\ref{eq:var2}) and in the third equality we used the linearity of the expectation.

%%%%%%%%%%%%%%%%%%%%%%%%%%%%%%%%%%%%%%%%%%
\subsection{Expectations of optical depth}
\label{subsec:tot}
%%%%%%%%%%%%%%%%%%%%%%%%%%%%%%%%%%%%%%%%%%
We compute the expectations with respect to the Gaussian-distributed optical depth to obtain the total expectation and variance in equations (\ref{eq:tot_I_exp}) and (\ref{eq:tot_I_var}). 
Since the optical depth always appears in an exponential, we are interested in,
\begin{equation}
\begin{split}
    \mathbb{E}_{\tau} \left[ e^{-\tau(z,s)} \right]
    \ &= \
    \int \D \tau(z,s) \ p\big(\tau(z,s)\big) \, e^{-\tau(z,s)} \\
    \ &= \
    \exp \left(-\tilde{\tau}(z,s) \ + \ \frac{1}{2} \, \tilde{\varepsilon}^{2}_{\tau}(z,s) \right) ,
\end{split}
\end{equation}
in which we used that the optical depth is Gaussian distributed, as in equation (\ref{eq:ptau}).
In the exponent, we recognise what we defined as the effective optical depth in equation (\ref{eq:optdepth_eff}).
Using the linearity of the expectation, this immediately yields the result in equation (\ref{eq:tot_I_exp}).

Similarly, for the variance, we also require,
\begin{equation}
\begin{split}
    \mathbb{E}_{\tau} \left[ e^{-\tau(z',s)} e^{-\tau(s',s)} \right]
    \ &= \ \mathbb{E}_{\tau} \left[ e^{-\tau(z',s)} \right] \mathbb{E}_{\tau} \left[ e^{-\tau(s',s)} \right] \\
    & \ \ \ \ \ + \ \text{Cov} \left[ e^{-\tau(z',s)}, \, e^{-\tau(s',s)} \right] .
\end{split}
\label{eq:?}
\end{equation}
Since the optical depth is Gaussian distributed, the exponential of minus the optical depth will follow a log-normal distribution.
The covariance of this log-normal distribution is given by,
\begin{equation}
\begin{split}
    & \text{Cov} \left[ e^{-\tau(z',s)}, \, e^{-\tau(s',s)} \right]
    \ = \ \\
    \ &  \ \Big(\exp\left(\text{Cov}\left[ -\tau(z',s), \, -\tau(s',s)  \right] \right) \ - \ 1 \Big) \\
    \ &  \ \times \exp \left(-\tilde{\tau}(z',s) - \tilde{\tau}(s',s) 
               \ + \ \frac{1}{2} \left(\tilde{\varepsilon}^{2}_{\tau}(z',s)  + \tilde{\varepsilon}^{2}_{\tau}(s',s) \right)
     \right) ,
\end{split}
\end{equation}
such that we can write the required expectation as,
\begin{equation}
\begin{split}
    & \mathbb{E}_{\tau} \left[ e^{-\tau(z',s)} e^{-\tau(s',s)} \right]
    \ = \ \\
    \ &  \ \exp\left(\text{Cov}\left[ -\tau(z',s), \, -\tau(s',s)  \right] \right) \\
    \ &  \ \times \exp \left(-\tilde{\tau}(z',s) - \tilde{\tau}(s',s) 
               \ + \ \frac{1}{2} \left(\tilde{\varepsilon}^{2}_{\tau}(z',s)  + \tilde{\varepsilon}^{2}_{\tau}(s',s) \right)
     \right) .
\end{split}
\label{eq:tar}
\end{equation}
The covariance of the optical depths can easily be derived from their joint Gaussian distribution, which yields,
\begin{equation}
\begin{split}
    & \text{Cov}\left[ -\tau(z',s), \, -\tau(s',s)  \right] \\
	\ &= \ \int_{z'}^{s} \D z''
	\int_{s'}^{s} \D s''
	\left(
	\kappa(s'',z'')
	-
	\kappa(\bs{a},s'')^{\T}
	\bm{K}_{\chi}^{-1}
    \kappa(\bs{a},z'')
	\right) \\
	\ &= \ \frac{1}{2} \left( \tilde{\varepsilon}^{2}_{\tau}(z',s) \ + \ \tilde{\varepsilon}^{2}_{\tau}(s',s) \ - \ \tilde{\varepsilon}^{2}_{\tau}(s',z')\right) ,
\end{split}
\end{equation}
where the last equality can be derived by subdividing the (2D) domain of integration.
Using the second effective optical depth variable (\ref{eq:optdepth_eff2}) to simplify notation, equation (\ref{eq:tar}) can be written as,
\begin{equation}
    \mathbb{E}_{\tau} \left[ e^{-\tau(z',s)} e^{-\tau(s',s)} \right]
    \ = \ e^{-\overline{\tau}(z',s)} e^{-\overline{\tau}(s',s)} e^{-\frac{1}{2}\tilde{\varepsilon}^{2}_{\tau}(s',z')} .
\end{equation}
Using the linearity of the expectation, we thus find,
\begin{equation}
\begin{split}
    \mathbb{E}_{\tau} \left[\tilde{\varepsilon}_{I}^{2} \right]
    \ &= \
	\int_{s_{0}}^{s} \D s'
	\int_{s_{0}}^{s} \D z'
	e^{-\overline{\tau}(s', \, s)}
	e^{-\overline{\tau}(z', \, s)}
	e^{-\frac{1}{2}\tilde{\varepsilon}^{2}_{\tau}(s',z')} \\
	& \ \ \ \ \ \times
	\left(
	\kappa(s',\,z')
	-
	\kappa(\bs{a},\,s')^{\T}
	\bm{K}_{\eta}^{-1}
    \kappa(\bs{a},\,z')
	\right) .
\end{split}
\label{eq:p1}
\end{equation}
Furthermore, using the same relations, we can find that,
\begin{equation}
\begin{split}
    \mathbb{E}_{\tau}\left[ \tilde{I}^{2} \right]
    \ = \ &
	\ I_{0}^{2} \, e^{-2\overline{\tau}(s_{0},s)} \\
	& \ + \
	2 \, I_{0} \, e^{-\overline{\tau}(s_{0},s)}
    \int_{s_{0}}^{s} \D s'
    H(s') \, e^{-\frac{1}{2}\tilde{\varepsilon}^{2}_{\tau}(s_{0},s')} \\
	& \ + \
	\int_{s_{0}}^{s} \D z'
    H(z')
    \int_{s_{0}}^{s} \D s'
    H(s')  \, e^{-\frac{1}{2}\tilde{\varepsilon}^{2}_{\tau}(z',s')}
\end{split}
\label{eq:p2}
\end{equation}
where we defined $H(s') \equiv \bs{\eta}^{\T} \, \bm{K}_{\eta}^{-1} \kappa(\bs{a}, s')\ e^{-\overline{\tau}(s',s)}$, to simplify notation.
In practice, equations (\ref{eq:p1}) and (\ref{eq:p2}) are difficult to work with due to their dependence on the cross term, $\tilde{\varepsilon}^{2}_{\tau}(z',s')$.
However, since this is a positive quantity, we can define an upper bound by removing it, such that equation (\ref{eq:p1}) simplifies to,
\begin{equation}
\begin{split}
    \mathbb{E}_{\tau} \left[\tilde{\varepsilon}_{I}^{2} \right]
    \ &\leq \
	\int_{s_{0}}^{s} \D s'
	\int_{s_{0}}^{s} \D z'
	e^{-\overline{\tau}(s', \, s)}
	e^{-\overline{\tau}(z', \, s)} \\
	& \ \ \ \ \ \times
	\left(
	\kappa(s',\,z')
	-
	\kappa(\bs{a},\,s')^{\T}
	\bm{K}_{\eta}^{-1}
    \kappa(\bs{a},\,z')
	\right) ,
\end{split}
\end{equation}
and, furthermore, equation (\ref{eq:p2}) simplifies to,
\begin{equation}
    \mathbb{E}_{\tau}\left[ \tilde{I}^{2} \right]
    \ \leq \
    \bar{I}^{2} .
\end{equation}
where, in analogy with equation (\ref{eq:tot_exp}), we defined,
\begin{equation}
	\bar{I} (s)
    \ \equiv \
	I_{0} \, e^{-\overline{\tau}(s_{0},s)}
	\ + \
	\bs{\eta}^{\T} \, \bm{K}_{\eta}^{-1}
    \int_{s_{0}}^{s} \D s' \
    \kappa(\bs{a}, s') \
    e^{-\overline{\tau}(s',s)}
\end{equation}
Combining all these results yields the practical inequality (\ref{eq:tot_I_var}).

%%%%%%%%%%%%%%%%%%%%%%%%%%%%%%%%%%%%%%%%%%%%%%%%%%

% Don't change these lines
\bsp	% typesetting comment
\label{lastpage}
\end{document}